\documentclass[10pt, journal , letterpaper]{IEEEtran}
\usepackage{amsmath,amssymb,amsfonts,amsthm}
\usepackage{romannum}
\usepackage{algorithm}
\usepackage{algorithmic}
\usepackage{graphicx,xcolor}
\usepackage{stfloats}
\usepackage[caption=false]{subfig}
\def\ie{{\it i.e.,\ \/}}
\def\eg{{\it e.g.,\ \/}}

\DeclareMathOperator{\tr}{tr}
\DeclareMathOperator{\diag}{diag}
\DeclareMathOperator{\blkdiag}{blkdiag}
\theoremstyle{definition}
\newtheorem{theorem}{Theorem}
\newtheorem{lemma}{Lemma}
\newtheorem{proposition}{Proposition}
\newtheorem*{remark}{Remark}
\makeatletter
\def\blfootnote{\gdef\@thefnmark{}\@footnotetext}
\makeatother
\def\Vc{\mathcal{V}}
\def\Cc{\mathcal{C}}

\def\Kc{\mathcal{K}}
\def\Pc{\mathcal{P}}
\def\Dc{\mathcal{D}}

\def\Mc{\mathcal{M}}
\def\ybf{\mathbf{y}}

\def\xbf{\mathbf{x}}
\def\Abf{\mathbf{A}}
\def\Bbf{\mathbf{B}}
\def\Hbf{\mathbf{H}}
\def\hbf{\mathbf{h}}
\def\gbf{\mathbf{g}}
\def\Wbf{\mathbf{W}}
\def\Ubf{\mathbf{U}}
\def\Vbf{\mathbf{V}}
\def\Dbf{\mathbf{D}}
\def\Ibf{\mathbf{I}}
\def\0bf{\mathbf{0}}
\def\1bf{\mathbf{1}}
\def\thetabf{\boldsymbol{\theta}}
\def\deltabf{\boldsymbol{\delta}}

\def\etabf{\boldsymbol{\eta}}

\begin{document}

\pagenumbering{arabic}


\title{Distributed Coordinated Precoding for \\ MIMO Cellular Network Virtualization }
\author{
        Juncheng Wang, \IEEEmembership{Student Member, IEEE},
        Min Dong, \IEEEmembership{Senior Member, IEEE},
        Ben Liang, \IEEEmembership{Fellow, IEEE},\\      
        Gary Boudreau, \IEEEmembership{Senior Member, IEEE},
        and Hatem Abou-zeid, \IEEEmembership{Member, IEEE}
\thanks{ J. Wang and B. Liang are with the  University of Toronto (e-mail: \{jcheng.wang, liang\}@ece.utoronto.ca). M. Dong is with the Ontario Tech University (e-mail: min.dong@ontariotechu.ca). G. Boudreau and H. Abou-zeid are with Ericsson Canada (e-mail: \{gary.boudreau, hatem.abou-zeid\}@ericsson.com). This work has been funded in part by Ericsson Canada and by the Natural
Sciences and Engineering Research Council (NSERC) of Canada.}
}

\maketitle


\begin{abstract}
This paper presents a new virtualization method for the downlink of a multi-cell multiple-input multiple-output (MIMO) network, to achieve service isolation among multiple Service Providers (SPs) that share the base station resources of an Infrastructure Provider (InP). Each SP designs a virtual precoder for its users in each cell, as its service demand to the InP, without the need to be aware of the existence of the other SPs or to know the channel state information (CSI) outside the cell. The InP performs network virtualization to meet the SPs' service demands while managing both the inter-SP and inter-cell interference. We consider coordinated multi-cell precoding at the InP and formulate an optimization problem to minimize a weighted sum of signal leakage and precoding deviation, with per-cell transmit power constraints. We propose a fully distributed semi-closed-form solution at each cell, without any CSI exchange across cells. We further propose a low-complexity scheme to allocate the virtual transmit power, for the InP to regulate between interference elimination and virtual demand maximization. Simulation results demonstrate that our precoding solution for network virtualization substantially outperforms the traditional spectrum isolation alternative. It can approach the performance of fully cooperative precoding when the number of antennas is large.
\end{abstract}



\section{Introduction}

High capital and operational expenses of wide-area wireless networks discourage wireless service providers (SPs) from technology upgrades and hinder new companies from entering the industry. As a solution to this, wireless network virtualization (WNV) has been proposed to reduce network deployment and operation expenses by abstracting and sharing physical resources \cite{WNV}-\cite{WNVsvy}. It decouples distinct parts of the network, making it easier for SPs to migrate to newer products and technologies. WNV is particularly important when physical infrastructure is expensive, such as a shopping mall with a high density of service requests but limited space to install many wireless base stations (BSs) from different SPs.

A virtualized network is generally composed of an infrastructure provider (InP), who owns and manages the physical infrastructure, and multiple SPs, who utilize the physical infrastructure to provide services to their subscribing users. For example, in existing commercial networks, the InP may refer to a mobile network operator, and the SPs may refer to mobile virtual network operators. The InP virtualizes the physical resources that it owns and splits them into virtual slices. The SPs lease these virtual slices and operate them to provide end-to-end services to their users without needing to know the underlying physical infrastructure or the existence of the other SPs. As a result, virtualization creates a set of logical entities from a given set of physical entities in a manner that is transparent to the SPs and their users, with the goal of \textit{service isolation} among SPs, \ie the service to the users of one SP is minimally affected by the other SPs.

For maximizing the potential of network virtualization, effective resource allocation is critical to ensure service isolation among SPs. However, service isolation is particularly challenging in wireless networks with the presence of interference \cite{sliceWNV}. Most existing works in the literature and commercially adopted systems apply \textit{strict} resource separation to achieve service isolation, by dividing the wireless spectrum, resource blocks, or antenna hardware among different SPs \cite{rp}\nocite{EE17}\nocite{V5G}\nocite{CRAN}\nocite{WNVNOMA}-\cite{Sgame}.
Such an approach is rooted in prior works on computer virtualization and
wired network virtualization, which has been shown to be highly effective.
However, as future networks adopt massive MIMO technology, strict resource separation limits the design space of virtualization
in the wireless environment, as it  does not explore the spatial dimension to allow more flexible wireless resource sharing to achieve higher power and spectral efficiency.

In contrast to strict resource separation, the wireless virtualization method proposed in \cite{Mpaper} leverages the interference suppression capability of massive multiple-input multiple-output (MIMO) when the InP is equipped with a large number of antennas. Service isolation via \textit{spatial} virtualization can be achieved through the precoding design at the InP. Instead of slicing resources physically, the InP can ensure the requirements of service isolation by using MIMO beamforming, while improving the overall network performance.

The existing works on spatial virtualization are limited to the single-cell case. In this work, we consider the virtualization design in a multi-cell MIMO network, where the InP-owned BS at each cell is simultaneously shared by multiple SPs to serve their subscribing users (in their respective virtual cells). Each SP designs the \textit{virtual} precoder as its virtualization demand based on the service needs and the local channel state information (CSI) of its users. The InP designs the \textit{actual} precoder with the goal to meet all SPs' demands, while ensuring service isolation among the SPs.

In a non-virtualized network, cooperative signal processing across the BSs of multiple cells has been identified as a key technique to mitigate inter-cell interference with significant performance improvement over the non-cooperative networks. Two levels of cooperation for transmitter precoding are often considered: \textit{cooperative} precoding \cite{COMP}-\cite{AsCOMP} and \textit{coordinated} precoding \cite{YuCOOD}\nocite{XWCOOD}-\cite{TonyCOOD}. The former refers to a fully cooperative scenario at the signal level, treating antennas at different BSs as distributed antennas forming a networked MIMO system. It requires data sharing among the BSs and stringent synchronization. In contrast, coordinated precoding does not require signal-level synchronization but only requires beamforming-level coordination without the need of data sharing.

In this work, we focus on the coordinated approach. Although multi-cell coordinated precoding has been extensively studied in non-virtualized wireless networks, new challenges arise in a  virtualized wireless network. Specifically, with service isolation, each SP provides its own desired precoding demand to the InP, and the InP designs the final precoder to meet each SP's demand. Oblivious to each other, each SP in a cell only has the CSI of its serving users (in its virtual cell), without access to the CSI of other SPs' users within the cell or  users in the other cells. It follows that the SPs' virtual precoding demands sent to the InP do not consider either inter-SP or inter-cell interference. As such, the InP must intelligently design the actual precoder to manage the interference among different SPs and cells, while trying to meet the SPs' virtual precoding demands. Thus, this virtualized coordinated precoding design problem is different from the traditional one and requires careful investigation for its solution.

The main contributions of this paper are summarized below:

\begin{itemize}

\item We design downlink WNV in a multi-cell MIMO system, by letting the InP decide the transmitter precoding to achieve service isolation among the SPs based on their individual virtual precoding demand as the service request. The design goal is to meet the SPs' service requests under interference management. To the best of our knowledge, this is the first work to design a virtualized multi-cell MIMO network with simultaneous utilization of all the antennas and channel resources, while managing both inter-SP and inter-cell interference.

\item We consider virtualization via coordinated precoding at the InP and formulate an optimization problem to minimize a  weighted sum of signal leakage and precoding deviation. We show that this problem can be decomposed into per-cell subproblems. This enables us to develop a fully distributed semi-closed-form solution at each cell, without any CSI exchange across cells. Our solution results in significant savings on the required computation and communication overhead. We also consider two other possible precoding optimization formulations with either signal leakage or precoding deviation as constraints, which are more complicated to solve. We show that they can be equivalently converted to the weighted sum cost minimization problem, for which we have a fully distributed semi-closed-form solution.
  
\item Since SPs are oblivious to each other, their virtual service demands (via virtual precoding) are absent of interference consideration. This requires the InP to carefully allocate the virtual transmit power for each SP's virtual service demand, to regulate between maximizing the SPs'  virtual service demands and managing interference. We propose a low-complexity virtual transmit power allocation scheme to control the trade-off between interference suppression and virtual demand maximization. With our proposed virtual transmit power, we show that the semi-closed-form precoding solution is further simplified to a closed form with minimal computational complexity.

\item We study the proposed precoding solution under the typical urban micro-cell Long-Term Evolution (LTE) network setting. Using both maximum ratio transmission (MRT) precoding and zero forcing (ZF) precoding as examples for the SPs' precoding choices, we show that our proposed precoding solution for network virtualization substantially outperforms the spectrum isolation alternative. In addition, it can approach the performance of a fully \textit{cooperative} network without service isolation among SPs, when the number of antennas becomes large such as in a massive MIMO system.
 
\end{itemize}

The rest of this paper is organized as follows. In Section \ref{Related Work},
we discuss the related work. In Section \ref{System Model}, we introduce the system model for network virtualization in a multi-cell MIMO system. In Section \ref{Single-Cell MIMO WNV}, we focus on the single-cell case and derive a semi-closed-form precoding solution and an effective virtual transmit power allocation scheme. In Section \ref{Multi-Cell MIMO WNV}, for the general multi-cell case, we discuss three coordinated precoding optimization formulations for virtualization, and present the proposed virtualized coordinated precoding solution and virtual transmit power allocation scheme. Simulation study and discussion are presented in Section \ref{Performance Evaluation}, followed by the conclusion in Section \ref{Conclusion}.

\textit{Notations}: The complex conjugate, Hermitian transpose, inverse, Moore-Penrose inverse, Frobenius norm, trace, and the $(i,j)$ element of a matrix $\mathbf{A}$ are denoted by $\mathbf{A}^*$, $\mathbf{A}^{H}$,
$\mathbf{A}^{-1}$, $\Abf^\dagger$, $\Vert \mathbf{A} \Vert_{F}$, $\tr\{\Abf\}$, and $[\mathbf{A}]_{i,j}$, respectively. The notation $\blkdiag\{\mathbf{A}_1,\dots,\mathbf{A}_n\}$
denotes a block diagonal matrix with diagonal elements being matrices $\mathbf{A}_1,\dots\mathbf{A}_n$, $\diag\{g_1,\dots,g_n\}$ denotes a diagonal matrix with diagonal elements
being $g_1,\dots,g_n$. Notation $\mathbf{0}$ denotes an all-zeros matrix,  $\mathbf{I}$ denotes an identity matrix, and $\mathbb{E}\{\cdot\}$ denotes expectation. For $\mathbf{g}$ being an $n\times1$ vector, $\mathbf{g}\sim\mathcal{CN}(\mathbf{0},\sigma^2\mathbf{I})$
means that $\mathbf{g}$ is a circular complex Gaussian random vector with
mean $\mathbf{0}$ and variance $\sigma^2\mathbf{I}$. 


\section{Related Work}
\label{Related Work}

WNV in MIMO systems has been studied mainly under two approaches in the literature. The first approach adopts strict physical resource isolation between the SPs \cite{rp}\nocite{EE17}\nocite{V5G}\nocite{CRAN}\nocite{WNVNOMA}-\cite{Sgame}. Among them, \cite{rp} and \cite{EE17} studied throughput maximization and energy minimization in orthogonal frequency division multiple access systems with massive MIMO. Sub-carriers were exclusively allocated to different SPs through a two-level hierarchical auction architecture in \cite{V5G}. Cloud radio networks and non-orthogonal multiple access techniques were combined with virtualized MIMO systems in \cite{CRAN} and \cite{WNVNOMA}. Antennas were assigned among the SPs through pricing for massive MIMO virtualization in \cite{Sgame}. However, restricting the SPs or even the users to orthogonal channels and exclusive subsets of antennas can lead to inefficient resource utilization and severe loss of system throughput compared with the complete sharing of all the antennas and channel resources.

The second approach uses MIMO precoding techniques to achieve spatial isolation among the SPs. Each SP utilizes all the antennas and channel resources, simultaneously with all other SPs,  and the InP uses signal processing techniques to manage the inter-SP interference \cite{Mpaper}, \cite{GLOBECOM19}\nocite{INFOCOM20}-\cite{SPAWC20}. However, the above works on spatial virtualization  are limited to the single-cell case. The spatial service isolation approach was first proposed in \cite{Mpaper}, where it was shown to substantially outperform the strict physical resource isolation approach. MIMO WNV in a fading environment was considered in \cite{GLOBECOM19} and \cite{INFOCOM20}, where online precoding schemes with perfect and imperfect CSI were proposed. A periodic precoder updating scheme was proposed for online MIMO precoding design for network virtualization with delayed CSI in \cite{SPAWC20}. Despite these works, MIMO precoding for network virtualization has not been investigated in a multi-cell system. In this work, we study service isolation via spatial virtualization in a multi-cell MIMO system. In this scenario, since each SP in a cell does not consider either the inter-SP interference within a cell or the inter-cell interference among the coordinated cells, it is challenging for the InP to manage the interference while meeting the service demands of SPs. To address this, we use a virtual transmit power to trade-off interference suppression and demand maximization. This strategy has not been considered in \cite{Mpaper}, \cite{GLOBECOM19}\nocite{INFOCOM20}-\cite{SPAWC20}.

For the traditional \textit{non-virtualized} cellular networks, multi-cell cooperative precoding  via multiple BSs at the signal level can effectively mitigate inter-cell interference and has been shown to significantly improve the system performance \cite{COMP}-\cite{AsCOMP}. However, the data streams of all users must be shared across all cooperating cells and the synchronization accuracy is critical. In contrast, multi-cell coordinated precoding only requires cooperation at the beamforming level without sharing the data streams \cite{YuCOOD}\nocite{XWCOOD}-\cite{TonyCOOD}. Weighted sum transmit power minimization subject to signal-to-interference-plus-noise ratio (SINR) constraints was studied in \cite{YuCOOD}. The joint power control and weighted sum rate maximization problem was addressed in \cite{XWCOOD}, where the proposed scheme requires CSI exchange across the coordinated cells. The problem of maximizing the minimum SINR subject to per-cell transmit power constraints was studied in \cite{TonyCOOD}, where the proposed scheme requires central update on the transmit power from each cell. Most existing coordinated precoding schemes for \textit{non-virtualized} networks are centralized and of high computational complexity and require CSI exchange across the coordinated cells  through the backhaul links or central update on the transmit power from each cell. It is desirable for practical systems to have a lower level of coordination, information exchange, and implementation complexity. Our general coordinated precoding solution for virtualized networks is fully distributed without any CSI exchange  across cells, and is in a semi-closed form. 

Besides the conventional cellular network architecture, cell-free massive MIMO has been recently proposed, where distributed single-antenna access points cooperatively transmit data to users \cite{HQNgo17}\nocite{ENayebi17}-\cite{EBjornson20}. The structure is a form of distributed MIMO, and it can be compared with a  co-located MIMO single-cell scenario. WNV is applicable in the cell-free structure, and our proposed spatial virtualization approach for WNV may be extended to cell-free massive MIMO systems. This is left for future research.
 

\section{System Model}
\label{System Model}

\begin{figure}[t]
\centering
\includegraphics[width=1\linewidth,trim=120 60 120 65,clip]{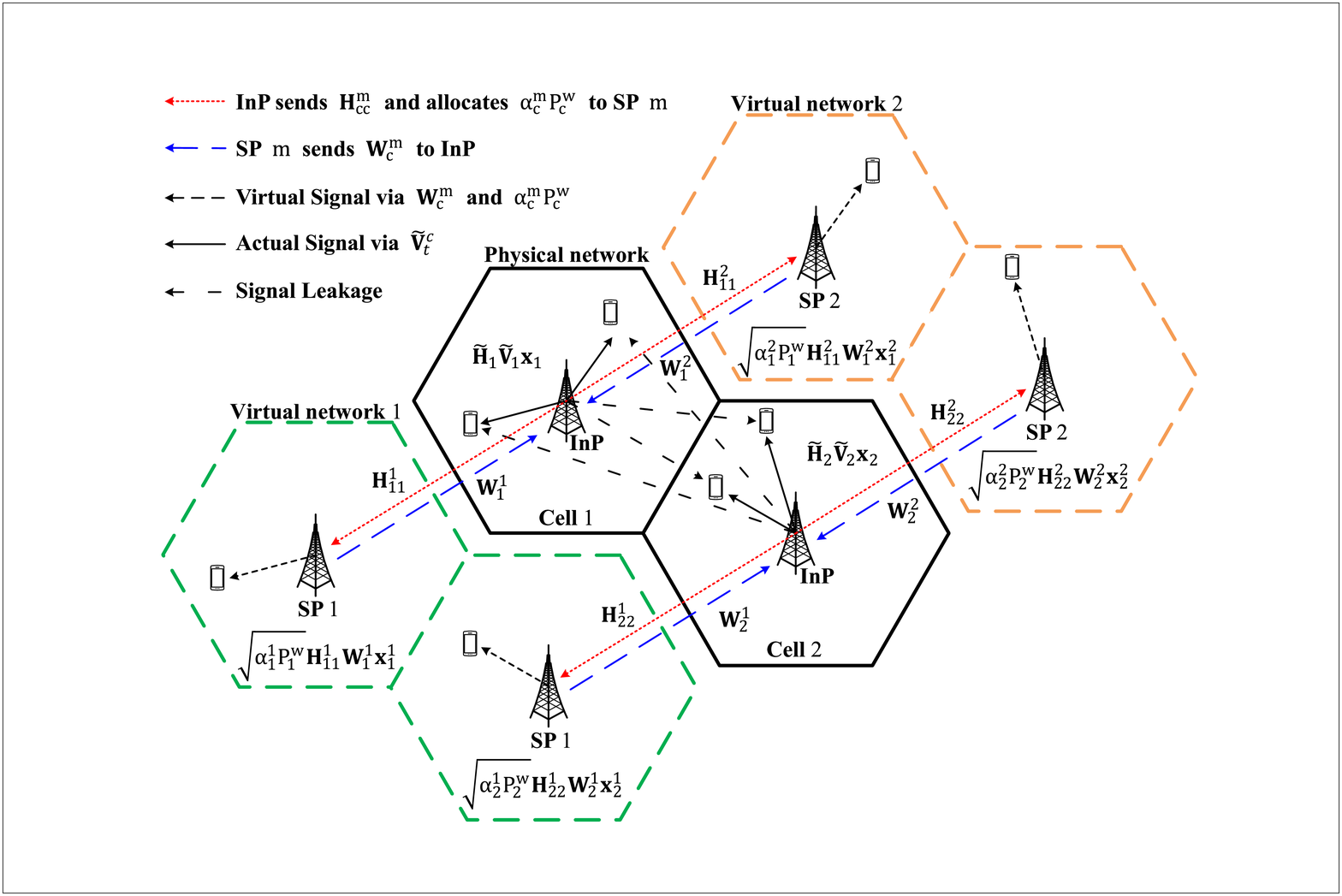}
\caption {An illustration of downlink coordinated MIMO network virtualization in a network with one InP and two SPs each serving its users in a virtual network.}
\label{Fig_WNV}
\vspace{-0mm}
\end{figure}

Consider a virtualized multi-cell downlink MIMO network in which an InP owns and operates the physical network infrastructure and multiple SPs are responsible for the services of their respective subscribing users. The InP performs cell virtualization at each cell for the SPs. The subscribing-user sets of different SPs are disjoint and each user is only served by its serving cell. To mitigate interference, multiple cells are coordinated at the transmission level, without CSI exchange across cells. An illustrative example is shown in Fig.~\ref{Fig_WNV}.

Specifically, we consider a total of $C$ cells owned by the InP. There are $M$ SPs that share the hardware, wireless spectrum, and transmission power provided by the InP at each cell BS. Let $\Cc=\{1,\dots,C\}$ and $\Mc=\{1,\dots,M\}$. The BS at each cell $c\in\mathcal{C}$ has $N_c$ antennas, so there is a total of $N=\sum_{c\in\Cc}N_c$ antennas in the network. Each SP $m\in\Mc$ has $K_c^m$ subscribing users in cell $c$. The total number of users in cell $c$ is $K_c=\sum_{m\in\Mc}K_c^{m}$, and that in the network is $K=\sum_{c\in\Cc}K_c$.

\subsection{Precoding Design by the InP and SPs}
\label{Sec:Precoding Design by InP and SPs}

Each SP designs its desirable precoding matrix for its users, and then sends it to the InP as its virtualization service demand. Specifically, let $\Hbf_{cl}^m\in\mathbb{C}^{K_c^m\times{N}_l}$ denote the channel of the $K_c^m$ users of SP $m$ in cell $c$ from the BS in cell $l$. In each cell $c$, the InP communicates with each SP $m$ the channel state $\Hbf_{cc}^m$ of SP $m$'s serving users in cell $c$. Based on the service needs and channel state $\Hbf_{cc}^m$ of its users, each SP $m$ designs a normalized precoding matrix $\Wbf_c^m\in\mathbb{C}^{N_c\times{K}_c^m}$
with $\Vert\Wbf_c^m\Vert_F^2=1$, to be sent to the InP as its precoding demand.  Note that each SP $m$ designs $\Wbf_c^m$ locally without knowledge of the other SPs' users in the  cell or the users in other cells, and it can choose any demanded precoding matrix. For illustration, in Section \ref{Performance Evaluation}, we will consider two most commonly used linear precoding schemes, \ie MRT precoding and ZF precoding.

Let $P_c^{\text{max}}$ denote the maximum transmit power at the BS in cell $c.$ After collecting the precoding demand $\Wbf_c^m$ from each SP $m$ in cell $c$, the InP allocates a \textit{virtual} transmit power $\alpha_c^m P_c^{\text{w}}$ to each SP $m$'s precoding demand, where $P_c^{\text{w}}\le P_c^{\text{max}}$ is the virtual transmit power allocated to cell $c$, and $\alpha_c^m$ is the virtual transmit power allocation factor for SP $m$ with $\sum_{m\in\mathcal{M}}\alpha_c^m =1$. Note that $\alpha_c^m$ indicates the fraction of InP's transmit power allocated to SP $m$ in cell $c$. We assume $\alpha_c^m$ is known apriori from the contractual agreement between SP $m$ and the InP. Its value may also depend on the priority of the SP, user density, some bidding mechanism, etc. Note that all existing spatial virtualization approaches assume that the InP allocates full transmit power to the SPs' precoding demands \cite{Mpaper}, \cite{GLOBECOM19}\nocite{INFOCOM20}-\cite{SPAWC20}.
This can lead to severe inter-SP and inter-cell interference, which in turn
deteriorates the system performance. In this work, we propose a more flexible
virtual transmit power allocation scheme at the InP to mitigate interference. It allows the InP to regulate between managing interference and maximizing each SP's demand. In Section \ref{Performance Evaluation}, we show that the proposed virtual transmit power allocation scheme substantially outperforms the full transmit power allocation approach.

Let $\xbf_c^m$ represent the downlink transmitted signal vector for the users of SP $m$ in cell $c$. With the precoding demand $\Wbf_c^m$ and virtual transmit power $\alpha_c^mP_c^{\text{w}}$, the \textit{virtual} received signal vector at the $K_c^m$ users of SP $m$ in cell $c$ is given by
\begin{align}
        \widetilde{\ybf}_{c}^m=\sqrt{\alpha_c^mP_c^{\text{w}}}\Hbf_{cc}^m\Wbf_c^m\xbf_{c}^m,\quad\forall{m}\in\mathcal{M}.\label{eq:tldyccm}
\end{align}
The virtual received signal vector  $\widetilde{\ybf}_{c}=[\hbox{$\widetilde{\ybf}_{c}^1$}^H,\dots,\hbox{$\widetilde{\mathbf{y}}_{c}^M$}^H]^H$
at all $K_c$ users in cell $c$ is given by
\begin{align}
        \widetilde{\ybf}_{c}=\sqrt{P_c^{\text{w}}}\Dbf_c\xbf_c,\quad\forall{c}\in\mathcal{C}\label{eq:tldycc}
\end{align}
where $\mathbf{x}_c=[{\mathbf{x}_c^1}^H,\dots,{\mathbf{x}_c^M}^H]^H$ is the overall signal vector for $K_c$ users in cell $c$ with $\mathbb{E}\{\mathbf{x}_c\mathbf{x}_c^H\}=\mathbf{I},\forall{c}\in\mathcal{C}$, and $\Dbf_c\triangleq\blkdiag\{\sqrt{\alpha_c^1}\Hbf_{cc}^1\Wbf_c^1,\dots,\sqrt{\alpha_c^M}\Hbf_{cc}^M\Wbf_c^M\}$
is the virtualization demand from cell $c$.

The InP virtualizes BS $c$ (and its serving cell) to meet the virtualization service demands of the SPs. Based on the channel states of all users, as well as the demanded precoding matrices $\Wbf_c^m$ from the SPs, the InP designs the actual downlink precoding $\widetilde{\Vbf}_c=[\Vbf_c^1,\dots,\Vbf_c^M]\in\mathbb{C}^{N_c\times{K}_c}$, to meet the SPs' demands, where $\Vbf_c^m\in\mathbb{C}^{N_c\times{K}_c^m}$ is the actual precoding designed for SP $m$ in cell $c$. The \textit{actual} received signal at the $K_c^m$ users originated from the serving BS using the InP-designed precoding matrix $\widetilde{\Vbf}_c$ at cell $c$ is given by
\begin{align}
        \ybf_{cc}^m=\Hbf_{cc}^m\Vbf_c^m\xbf_c^m+\sum_{i\neq{m},i\in\Mc}\Hbf_{cc}^m\Vbf_c^i\xbf_c^i,\quad\forall{m}\in\mathcal{M}\label{eq:yccm}
\end{align}
where the second term is the intra-cell inter-SP interference to the users of SP $m$ from the other SPs. Note that $\ybf_{cc}^m$ only contains signals from the BS in cell $c$ and does not contain inter-cell interference. The actual received signal at users in cell $l$ from the BS in cell $c$ is given by
\begin{align}
        \ybf_{lc}=\bar{\Hbf}_{lc}\widetilde{\Vbf}_c\xbf_c,\quad\forall{l,c}\in\mathcal{C}.\label{eq:ycc}
\end{align}
where $\bar{\Hbf}_{lc}=[{\Hbf_{lc}^1}^H,\dots,{\Hbf_{lc}^M}^H]^H\in\mathbb{C}^{K_l\times{N}_c}$
is the channel state between the $K_l$ users in cell $l$ and the BS in cell $c$.
 
As shown in Fig.~\ref{Fig_WNV}, the virtualization procedure in each cell $c$ is summarized as follows: 1)~the InP communicates the local channel state $\Hbf_{cc}^m$ of subscribing users to each SP $m$; 2)~SP~$m$ designs the normalized virtual precoding matrix $\Wbf_c^m$ and sends it to the InP as the virtualization service demand; 3)~the InP allocates a virtual transmit power $\alpha_c^mP_c^{\text{w}}$ to each SP $m$, and designs the actual precoding matrix $\widetilde{\Vbf}_c$ for downlink transmission for users in cell $c$.\footnote{Note
that in our precoding design, each SP $m$ designs $\mathbf{W}_c^m$ locally and does not handle intra-SP inter-cell interference. The inter-cell interference is solely handled by the InP.}
 
\subsection{Signal Leakage and Precoding Deviation}
\label{Sec:Signal Leakage and Precoding Deviation}

Since $\Wbf_c^m$ is designed locally by SP $m$ without considering either inter-SP or inter-cell interference, the InP needs to design the actual precoding $\widetilde{\Vbf}_c$ to mitigate interference and ensure the actual received signal $\ybf_{cc}^m$ in (\ref{eq:yccm}) reflects the service demand of SP $m$ in cell $c$. For this purpose, we consider two design metrics. First, to quantity the difference between the actual precoding by the InP and the virtual precoding by the SPs for cell $c$, we define the \textit{precoding deviation} based on (\ref{eq:tldycc}) and (\ref{eq:ycc}) as 
\begin{align}
       \!\!\rho_c(\widetilde{\Vbf}_c)\triangleq\mathbb{E}_{\xbf_c}\{\Vert\ybf_{cc}-\widetilde{\ybf}_{c}\Vert_F^2\}=\!\Vert\bar{\Hbf}_{cc}\widetilde{\Vbf}_c-\sqrt{P_c^{\text{w}}}\Dbf_c\Vert_F^2.\!\label{EQ:MC_rho}
\end{align}
Note that the precoding deviation defined above serves as a natural performance metric to quantify how well the SPs' service demands are satisfied by the InP in the virtualized network with service isolation. It is a part of the unique demand-response mechanism between the SPs and the InP.

Next, to quantify inter-cell interference by the InP precoding, we consider the \textit{signal leakage} defined as
\begin{align}
        f_c(\widetilde{\Vbf}_c)\triangleq\mathbb{E}_{\xbf_c}\left\{\sum_{l\neq{c},l\in\Cc}\Vert\ybf_{lc}\Vert_F^2\right\}=\sum_{l\neq{c},l\in\Cc}\Vert\bar{\Hbf}_{lc}\widetilde{\Vbf}_c\Vert_F^2.\label{EQ:MC_f}
\end{align}
It indicates the amount of inter-cell interference generated by cell $c$ to all the other cells. The signal leakage is often considered as a design criterion for interference management in conventional non-virtualized MIMO systems \cite{SLNR}.

Ideally, the InP designs the precoding matrix $\widetilde{\Vbf}_c$ to eliminate inter-SP interference in cell $c$ and inter-cell interference, such that it meets the precoding demands with zero precoding deviation $\rho_c(\widetilde{\Vbf}_c)=0$ and generates no signal leakage to other cells $f_c(\widetilde{\Vbf}_c)=0$. However, these two cannot be satisfied in general. This is because interference management limits the degrees of freedom for precoding within an SP's user set. In Section \ref{Single-Cell MIMO WNV}, we first consider a single-cell MIMO virtualization design to minimize the precoding deviation. We then extend the virtualization design to the multi-cell scenario where we consider the trade-off between precoding deviation and signal leakage. This trade-off is unique to the virtualization design and has not been considered in traditional precoding problems before.

\section{Single-Cell MIMO  Network Virtualization}
\label{Single-Cell MIMO WNV}

For clarity of presentation, we first consider network virtualization design in a single-cell MIMO system. The results obtained in the single-cell case will be used for the multi-cell case in Section \ref{Multi-Cell MIMO WNV}. We note that the proposed solution here is different from those in \cite{Mpaper}, \cite{GLOBECOM19}\nocite{INFOCOM20}-\cite{SPAWC20}. The objective of \cite{Mpaper} is power minimization, while \cite{GLOBECOM19}\nocite{INFOCOM20}-\cite{SPAWC20} focus on online optimization only. Furthermore, none of these works consider virtual transmit power.

\subsection{Precoding Deviation Minimization}

Consider MIMO virtualization in a single cell. Since there is only one cell, to ease the description, we simplify the notations to omit the cell index $c$. Specifically, the BS has $N$ antennas. Each SP $m$ has $K^m$ users, and their channels from the BS is $\Hbf^m\in{\mathbb{C}}^{K^m\times{N}}$. Based on the virtualization procedure described in Section \ref{Sec:Precoding Design by InP and SPs}, SP $m$'s service demand is the normalized precoding matrix $\Wbf^m\in\mathbb{C}^{N\times{K}^m}$. The virtual transmit power that the InP allocates to SP $m$ is $\alpha^mP^\text{w}$, where $P^\text{w}$ and $\alpha^m$ are as defined in Section \ref{Sec:Precoding Design by InP and SPs}, with cell index $c$ removed. The global channel state of users of all SPs in the cell is denoted by $\Hbf=[{\Hbf^1}^H,\dots,{\Hbf^M}^H]^H\in\mathbb{C}^{K\times{N}}$.
The InP designs the precoding matrix $\Vbf=[\Vbf^1,\dots,\Vbf^M]\in\mathbb{C}^{N\times{K}}$, where $\Vbf^m\in\mathbb{C}^{N\times{K}^m}$ corresponds to the precoding for the users of SP $m$.

Following (\ref{eq:tldyccm}) and (\ref{eq:yccm}), the virtual received signal based on the service needs of SP $m$ is given by
\begin{align*}
        \widetilde{\ybf}^m=\sqrt{\alpha^m{P}^\text{w}}\Hbf^m\Wbf^m\xbf^m,\quad\forall{m}\in\mathcal{M}
\end{align*}
where $\xbf^m$ is the downlink messages for the $K^m$ users of SP $m$, and the actual received signal at the users of SP $m$ is given by
\begin{align*}
        \ybf^m=\Hbf^m\Vbf^m\xbf^m+\sum_{i\neq{m},i\in\Mc}\Hbf^m\Vbf^i\xbf^i,\quad\forall{m}\in\mathcal{M}.
\end{align*}
Based on (\ref{EQ:MC_rho}), the \textit{precoding deviation} between the actual precoding by the InP and the virtual precoding demand by the SPs is given by
\begin{align}
        \rho(\Vbf)\triangleq\Vert\Hbf\Vbf-\sqrt{P^\text{w}}\Dbf\Vert_F^2\label{EQ:SC_rho}
\end{align}
where $\Dbf\triangleq\blkdiag\{\sqrt{\alpha^1}\Hbf^1\Wbf^1,\dots,\sqrt{\alpha^M}\Hbf^M\Wbf^M\}$. Recall that the virtual transmit power $P^\text{w}$ regulates between interference suppression and each SP's demand maximization. Since it is a single-cell scenario, the signal leakage is not considered.

For the virtualized MIMO system, our goal for the InP precoding design is to minimize the the precoding deviation subject to the maximum transmit power limit
\begin{align}
        \Pc:\quad\min_{\Vbf}\quad&\rho(\Vbf)\notag\\
        \text{s.t.}~~~&\Vert\Vbf\Vert_F^2-P^{\text{max}}\le0.\label{EQ:SC_g}
\end{align}
 Note that the virtual transmit power $P^\text{w}$ serves as a tuning parameter in $\rho(\Vbf)$ to reach a certain desired system performance for SPs (\eg minimum rate, sum-rate). Next, we show that the problem of precoding deviation minimization $\Pc$ leads to an interesting semi-closed-form solution.

\subsection{Semi-Closed-Form Precoding Solution}
\label{SEC:Semi-Closed-Form Precoding Solution}

Now we solve the precoding deviation minimization problem $\Pc$
to obtain the optimal solution $\Vbf^\circ$ at the InP for any given virtual
transmit power $P^\text{w}\le{P}^{\text{max}}$. Note that $\Pc$ is a convex problem and in fact is a constrained least-square problem. We can derive a semi-closed-form solution using the Karush-Kuhn-Tucker (KKT) conditions \cite{Boyd}. 

The Lagrangian for $\Pc$ is 
\begin{align}
        L(\Vbf,\lambda)=\Vert\Hbf\Vbf-\sqrt{P^\text{w}}\Dbf\Vert_F^2+\lambda(\Vert\Vbf\Vert_F^2-P^{\text{max}})\label{EQ:SC_L}
\end{align}
where $\lambda$ is the Lagrange multiplier for the power constraint (\ref{EQ:SC_g}). The KKT conditions for $(\Vbf^\circ,\lambda^\circ)$ being globally optimal are given by
\begin{align}
        \nabla{L}(\Vbf^\circ,\lambda^\circ)=\Hbf^H(\Hbf\Vbf^\circ&-\sqrt{P^\text{w}}\Dbf)+\lambda^\circ\Vbf^\circ=\mathbf{0},\label{EQ:SC_KKT1}\\
        \Vert\Vbf^\circ\Vert_F^2&\le P^{\text{max}},\label{EQ:SC_KKT2}\\
        \lambda^\circ&\ge0,\label{EQ:SC_KKT3}\\
        \lambda^\circ(\Vert\Vbf^\circ\Vert_F^2&-P^{\text{max}})=0\label{EQ:SC_KKT4}
\end{align}
where in (\ref{EQ:SC_KKT1}), we use the equalities $\Vert\Abf\Vert_F^2=\tr\{\Abf\Abf^H\}$, $\nabla_{\Bbf^*}\tr\{\Abf\Bbf^H\}=\Abf$, and $\nabla_{\Bbf^*}\tr\{\Abf\Bbf\}=\0bf$ \cite{CGD} to derive the partial derivative of $L(\Vbf^\circ,\lambda^\circ)$ with respect to the complex conjugate of $\Vbf^\circ$.

Based on (\ref{EQ:SC_KKT1})-(\ref{EQ:SC_KKT4}), we discuss the optimal solution in the following two cases.

\subsubsection{$\lambda^\circ=0$} 
From (\ref{EQ:SC_KKT1}), the optimal solution must satisfy
\begin{align}
        \Hbf^H\Hbf\Vbf^\circ&=\sqrt{P^\text{w}}\Hbf^H\Dbf.\label{EQ:SC_HVD}
\end{align}
The solution $\Vbf^\circ$ depends on the relation of $N$ and $K$, given in the following two subcases. \romannum{1}) $N\ge{K}$: In this case, $\Hbf^H\Hbf\in\mathbb{C}^{N\times{N}}$ is rank deficient, and there are infinitely many solutions for $\Vbf^\circ$. We choose $\Vbf^\circ$ to minimize $\Vert\Vbf^\circ\Vert_F^2$ subject to (\ref{EQ:SC_HVD}), which is an under-determined least square problem with a closed-form solution given by
\begin{align}
        \Vbf^\circ=\sqrt{P^\text{w}}\Hbf^H(\Hbf\Hbf^H)^{-1}\Dbf.\label{EQ:SC_OPT1}
\end{align}
Note that in the special case $N=K$, (\ref{EQ:SC_OPT1}) can be simply written as $\Vbf^\circ=\sqrt{P^\text{w}}\Hbf^{-1}\Dbf$. \romannum{2}) $N<K$: In this case, $\Hbf^H\Hbf\in\mathbb{C}^{N\times{N}}$ is full rank\footnote{For users
at different locations, it is typically satisfied that the channels from the BS to users are linearly independent, \ie $\Hbf$ is of full rank.}, and we have a unique solution for $\Vbf^\circ$ given by
\begin{align}
        \Vbf^\circ=\sqrt{P^\text{w}}(\Hbf^H\Hbf)^{-1}\Hbf^H\Dbf.\label{EQ:SC_OPT2}
\end{align}
For both subcases \romannum{1}) and \romannum{2}), $\Vbf^\circ$ in (\ref{EQ:SC_OPT1}) or (\ref{EQ:SC_OPT2}) is optimal only if it satisfies the power constraint (\ref{EQ:SC_KKT2}). Otherwise, it means the condition $\lambda^\circ=0$ for in Case 1) does not hold at optimality, and we have $\lambda^\circ>0$, which is discussed in the next case.

\subsubsection{$\lambda^\circ>0$} From (\ref{EQ:SC_KKT1}), we have
\begin{align}
        \Vbf^\circ=\sqrt{P^\text{w}}(\Hbf^H\Hbf+\lambda^\circ\Ibf)^{-1}\Hbf^H\Dbf\label{EQ:SC_OPT3}
\end{align}
where by (\ref{EQ:SC_KKT4}), $\lambda^\circ$ is such that $P^\text{w}\Vert(\Hbf^H\Hbf+\lambda^\circ\Ibf)^{-1}\Hbf^H\Dbf\Vert_F^2=P^{\text{max}}$.
The optimal $\lambda^\circ>0$ can be obtained using the bisection search. The search range is described in the following proposition. 

\begin{proposition}\label{LM:Bsec}
For $\Vbf^\circ$ in (\ref{EQ:SC_OPT3}), the optimal Lagrange multiplier $\lambda^\circ$ lies in the interval $\lambda^\circ\in\left(0,\Vert\Hbf\Vert_F^2\sqrt{\frac{NP^\text{w}}{P^{\text{max}}}}\right]$.
\end{proposition}
\textit{Proof:} See Appendix \ref{APP:Bsec}.

The optimal solution $\Vbf^\circ$ for $\Pc$ is the one that results in the minimum $\rho(\Vbf)$ in $\Pc$. Note that if $\lambda^\circ=0$ at optimality, we have a closed-form solution for $\Vbf^\circ$ in (\ref{EQ:SC_OPT1}) or (\ref{EQ:SC_OPT2}). Otherwise, we have a semi-closed form solution for $\Vbf^\circ$ in (\ref{EQ:SC_OPT3}), where $\lambda^\circ>0$ can be obtained by the bisection search within the interval shown in Proposition \ref{LM:Bsec}. The computational complexity for calculating $\Vbf^\circ$ is dominated by matrix inversion, and thus is $\mathcal{O}(\min\{N,K\}^3)$. 

\begin{remark}
Note that our semi-closed-form solution structure in (\ref{EQ:SC_OPT3}) is similar to transmit minimum-mean-square-error (MMSE) precoding. However, there are some key differences between the two: 1) the solution in (\ref{EQ:SC_OPT3}) contains an additional matrix $\mathbf{D}$ that represents the virtualization demand of the SPs; 2) the solution in (\ref{EQ:SC_OPT3}) contains a  virtual transmit power $P^{\text{w}}$ to regulate interference suppression and demand maximization.
\end{remark}

\subsection{Virtual Transmit Power Allocation $P^{\normalfont\text{w}}$}
\label{SEC:Virtual Transmit Power Allocation}

Recall that $\Wbf^m$ is a normalized precoding matrix as SP~$m$'s service demand. It indicates the \emph{relative} desired service that the SP provides among its users. Proper power allocation is required to reflect the actual desired service quality (\eg rate). Since the InP needs to mitigate inter-SP interference (which uses some power), the transmit power allocated to each SP for its own precoding purpose is less than the maximum transmit power $P^{\text{max}}$. The virtual transmit power $P^\text{w}$ in $\rho(\Vbf)$ is intended to regulate between interference suppression and each SP's demand maximization. However, the optimization of the system performance (\eg  minimum rate, sum rate) w.r.t. $P^\text{w}$ is usually non-convex. Also, the range to search the optimal $P^\text{w}$ could be very large, making the search computationally expensive. Therefore, we propose an intuitive virtual transmit power allocation scheme to simplify the searching process. We will show in Section \ref{Performance Evaluation} that  the proposed virtual transmit power allocation strategy achieves system performance that is close to the optimum. To the best of our knowledge, the use of virtual transmit power to trade-off between interference suppression and virtualization
demand maximization has not been considered in the existing literature.

Consider the idealized case where the actual precoding matrix $\Vbf$ achieves zero precoding deviation $\rho(\Vbf)=0$, \ie
\begin{align}
        \Hbf\Vbf-\sqrt{P^\text{w}}\Dbf=\0bf\label{EQ:SC_0Dev}
\end{align}
while meeting the power constraint in (\ref{EQ:SC_g}). We notice that the virtual transmit power $P^{\text{w}}$ can be viewed as a \emph{power regularization} factor for the least-square precoding solution $\Vbf^\circ$ in (\ref{EQ:SC_OPT1}) or (\ref{EQ:SC_OPT2}), such that $\Vert\Vbf^\circ\Vert_F^2\le{P}^{\text{max}}$ in (\ref{EQ:SC_g}). It follows that the maximum value of $P^{\text{w}}$ for $\Vbf^\circ$ in (\ref{EQ:SC_OPT1}) or (\ref{EQ:SC_OPT2}) to satisfy $(\ref{EQ:SC_g})$ with equality is given by
\begin{align}
        P^{\text{w}\circ}=\min\left\{\frac{P^{\text{max}}}{\Vert \Hbf^\dagger\Dbf\Vert_F^2},P^{\text{max}}\right\}.\label{EQ:SC_Pw}
\end{align}
Note that under the precoding matrix $\Vbf^\circ$ in (\ref{EQ:SC_OPT1}) or (\ref{EQ:SC_OPT2}), the SINR of each user in the cell monotonically increases with $P^{\text{w}}$, and thus is maximized under the virtual transmit power $P^{\text{w}\circ}$ in (\ref{EQ:SC_Pw}). As a result, we propose to use $P^{\text{w}\circ}$ in the solution to (\ref{EQ:SC_0Dev}) $\Vbf^\circ=\sqrt{P^{\text{w}\circ}}\Hbf^\dagger\Dbf$.

As indicated earlier, when $N\ge{K}$, the solution to (\ref{EQ:SC_0Dev}) is not unique. The optimal $\Vbf^\circ$ that minimizes $\Vert\Vbf\Vert_F^2$ is given in (\ref{EQ:SC_OPT1}), where $\Hbf^\dagger=\Hbf^H(\Hbf\Hbf^H)^{-1}$. In this case, the precoding solution $\Vbf^\circ$ completely nulls the inter-SP interference, which is desired for the service isolation among SPs in WNV, and  $P^{\text{w}\circ}$ is the maximum possible power for each SP's demand after inter-SP interference cancellation. When $N<K$, the equation in (\ref{EQ:SC_0Dev}) is over-determined and the optimal $\Vbf^\circ$ is the least-square solution for $\Vert\Hbf\Vbf-\sqrt{P^{\text{w}}}\Dbf\Vert_F^2$ given in (\ref{EQ:SC_OPT2}), where $\Hbf^\dagger=(\Hbf^H\Hbf)^{-1}\Hbf^H$. In this case, since the inter-SP interference cannot be completely eliminated, the virtual transmit power $P^{\text{w}\circ}$ regularizes the interference suppression and the demand maximization at SPs.

Note that when the virtual transmit power $P^{\text{w}\circ}$ in (\ref{EQ:SC_Pw}) is used, we effectively adopt the closed-form optimal solution $\mathbf{V}^\circ$ to $\mathcal{P}$ given in (\ref{EQ:SC_OPT1}) or (\ref{EQ:SC_OPT2}), instead of the semi-closed-form solution in (\ref{EQ:SC_OPT3}), which is invoked only when $P^\text{w}>P^{\text{w}\circ}$. Through simulation, we will show that this choice of $P^{\text{w}\circ}$ (and $\mathbf{V}^\circ$) results in system performance at each SP (\eg average rate or minimum rate) close to the maximum, and thus is a near optimal value. 

\section{Multi-Cell MIMO Network Virtualization}
\label{Multi-Cell MIMO WNV}

In this section, we extend the MIMO precoding virtualization solution of the single-cell case to the multi-cell scenario. For a multi-cell MIMO network, the level of coordination and how to perform distributed implementation are two critical issues. We consider multi-cell precoding coordination in the MIMO WNV systems. Our proposed coordinated precoding scheme for network virtualization naturally leads to a \textit{fully distributed} implementation at each cell.

\subsection{Precoding Virtualization Formulation}

In a virtualized multi-cell MIMO system, due to inter-cell interference, the leakage $f_c(\widetilde{\Vbf}_c)$ in (\ref{EQ:MC_f}) and the precoding deviation $\rho_c(\widetilde{\Vbf}_c)$ in (\ref{EQ:MC_rho}) cannot be completely eliminated. In general, the two criteria constrain each other in the design. As a result, the system performance (\eg minimum rate, sum rate) depends on both $f_c(\widetilde{\Vbf}_c)$ and $\rho_c(\widetilde{\Vbf}_c)$. We design the InP precoding to trade-off the effect of signal leakage and precoding deviation to achieve certain desired system performance.

For the precoding virtualization design at the InP, we consider three problem formulations as follows:

\subsubsection{Weighted leakage and precoding deviation}

We first consider the precoding optimization at the InP to minimize a weighted sum of signal leakage and precoding deviation, subject to the per-cell maximum transmit power constraints, given by
\begin{align}
        \Pc^{\text{w}}(\thetabf):\quad\min_{\{\widetilde{\Vbf}_c\}}\quad&\sum_{c\in\Cc}(1-\theta_c)f_c(\widetilde{\Vbf}_c)+\theta_c\rho_c(\widetilde{\Vbf}_c)\notag\\
        \text{s.t.}~~~&\Vert\widetilde{\Vbf}_c\Vert_F^2-P_c^{\text{max}}\le0,\quad\forall{c}\in\Cc\label{eq:Pw_gc}
\end{align}
where $\thetabf=[\theta_1,\dots,\theta_C]^T$, with $\theta_c\in[0,1]$, is the weight vector that sets the relative importance between the signal leakage and precoding deviation in the cost function; it can be tuned by the InP to optimize certain specified system performance for each SP.

\subsubsection{Constrained leakage minimization}

We can also formulate the problem to minimize the signal leakage, while limiting the precoding deviation below a threshold. The resulting constrained leakage minimization problem is given by
\begin{align*}
        \Pc^{\text{lk}}(\deltabf):\quad\min_{\{\widetilde{\Vbf}_c\}}\quad&\sum_{c\in\Cc}f_c(\widetilde{\Vbf}_c)\notag\\
        \text{s.t.}~~~ & \rho_c(\widetilde{\Vbf}_c)\le\delta_c \text{~and~}(\ref{eq:Pw_gc}),\quad\forall{c\in\Cc},
\end{align*}
where $\deltabf=[\delta_1,\dots,\delta_C]^T$, with $\delta_c\in[0,\infty)$, is the limit on the precoding deviation that can be tuned by the InP.

\subsubsection{Constrained precoding deviation minimization}

The inverse problem for $\Pc^{\text{lk}}(\deltabf)$ is to minimize the precoding deviation, subject to the signal leakage constraint. The resulting constrained precoding deviation minimization problem is given by
\begin{align*}
        \Pc^{\text{d}}(\etabf):\quad\min_{\{\widetilde{\Vbf}_c\}}\quad&\sum_{c\in\Cc}\rho_c(\widetilde{\Vbf}_c)\notag\\
        \text{s.t.}~~~ & f_c(\widetilde{\Vbf}_c)\le\eta_c\text{~and~}(\ref{eq:Pw_gc}),\quad\forall{c\in\Cc},
\end{align*}
where $\etabf=[\eta_1,\dots,\eta_C]^T$, with $\eta_c\in[0,\infty)$, is the limit imposed on the signal leakage. 

Note that the above three problems $\Pc^{\text{w}}(\thetabf)$, $\Pc^{\text{lk}}(\deltabf)$, and $\Pc^{\text{d}}(\etabf)$ are all convex. Furthermore, $\Pc^{\text{lk}}(\deltabf)$ and $\Pc^{\text{d}}(\etabf)$ can be subsumed by $\Pc^{\text{w}}(\thetabf)$. In the following, we discuss the relation of $\Pc^{\text{lk}}(\deltabf)$ to $\Pc^{\text{w}}(\thetabf)$ as an example.

First, we note that $\Pc^{\text{w}}(\thetabf)$ can be  decomposed into $C$ subproblems, each corresponding to a local precoding design optimization problem for cell $c$, given by
\begin{align}
        \Pc^{\text{w}}_c(\theta_c):\quad\min_{\widetilde{\Vbf}_c}
        \quad & (1-\theta_c)f_c(\widetilde{\Vbf}_c)+\theta_c\rho_c(\widetilde{\Vbf}_c)\notag\\
        \text{s.t.}~~~&\Vert\widetilde{\Vbf}_c\Vert_F^2-P_c^{\text{max}}\le0.\label{EQ:MC_gc}
\end{align}
Note that in $\Pc^{\text{w}}_c(\theta_c)$, the objective is a locally weighted sum of leakage and precoding deviation, which only depends on the local channel states $\{\bar{\Hbf}_{lc}\}_{l=1}^C$. As a result, the InP designs $\widetilde{\Vbf}_c$ based only on $\{\bar{\Hbf}_{lc}\}_{l=1}^C$ to minimize the local objective in cell $c$, subject to the maximum transmit power constraint. As such, the coordinated precoding optimization problem in $\Pc^{\text{w}}(\thetabf)$ is fully distributed, without any CSI exchange across cells or cental update on transmit power from each cell.

The problem $\Pc^{\text{lk}}(\deltabf)$ can also be decomposed into $C$ subproblems,
each being a local precoding optimization problem at cell $c$, given by
\begin{align*}
        \Pc^{\text{lk}}_c(\delta_c):\quad\min_{\widetilde{\Vbf}_c}\quad&f_c(\widetilde{\Vbf}_c)\notag\\
        \text{s.t.}~~~ & \rho_c(\widetilde{\Vbf}_c)\le\delta_c\text{~and~}(\ref{EQ:MC_gc}).
\end{align*}

However, there is always a feasible solution to $\Pc_c^{\text{w}}(\theta_c)$, while $\Pc_c^{\text{lk}}(\delta_c)$ has a feasibility issue depending on the value of $\delta_c$. Let $\widetilde{\Vbf}_c^{\text{w}\circ}(\theta_c)$
denote an optimal solution to $\Pc_c^{\text{w}}(\theta_c)$, the following lemma gives a necessary and sufficient condition on the feasibility of $\Pc_c^{\text{lk}}(\delta_c)$.

\begin{lemma}\label{LM:fea}
Problem $\Pc^{\text{lk}}_c(\delta_c)$ is feasible if and only if
\begin{align}
        \delta_c\ge\delta_c^{\text{w}}\triangleq\rho_c(\widetilde{\Vbf}_c^{\text{w}\circ}(1)).\label{EQ:defdeltac-}
\end{align}
\end{lemma}
\textit{Proof:} See Appendix \ref{APP:fea}.

In Lemma \ref{LM:fea}, the feasibility region of $\Pc^{\text{lk}}_c(\delta_c)$ is shown in terms of the precoding deviation limit $\delta_c^{\text{w}}$. This limit depends on the maximum transmit power $P_c^{\text{max}}$, virtual transmit power $P_c^{\text{w}}$, and the CSI of all users in cell $c$.

The following lemma gives the condition on $\delta_c$ such that the strong duality holds for $\Pc_c^{\text{lk}}(\delta_c)$.

\begin{lemma}\label{LM:dual}
The strong duality holds for $\Pc_c^{\text{lk}} (\delta_c)$, for $\delta_c>\delta_c^{\text{w}}$.
\end{lemma}
\textit{Proof:} See Appendix \ref{APP:dual}.

By Lemma \ref{LM:dual}, for any $\delta_c>\delta_c^{\text{w}}$, we can solve $\Pc_c^{\text{lk}}(\delta_c)$ through its dual problem instead. The Lagrange function for $\Pc_c^{\text{lk}}(\delta_c)$ is given by
\begin{align*}
        &L_c^{\text{lk}}(\widetilde{\Vbf}_c,\nu_c,\mu_c;\delta_c)\\
        &=f_c(\widetilde{\Vbf}_c)+\nu_c[\rho_c(\widetilde{\Vbf}_c)-\delta_c]+\mu_c(\Vert\widetilde{\Vbf}_c\Vert_F^2-P_c^{\text{max}})
\end{align*}
where $\nu_c\ge0$ and $\mu_c\ge0$ are the Lagrange\ multipliers associated with the precoding deviation constraint and the maximum transmit power constraint, respectively. The dual problem of $\Pc_c^{\text{lk}} (\delta_c )$ is given by
\begin{align*}
       \Dc_c^{\text{lk}}(\delta_c):\quad\max_{\nu_c\ge 0,\mu_c\ge0}\min_{\widetilde{\Vbf}_c}\quad{L}_c^{\text{lk}}(\widetilde{\Vbf}_c,\nu_c,\mu_c;\delta_c).
\end{align*}
Let $(\widetilde{\Vbf}_c^{\text{lk}\circ}(\delta_c),\nu_c^\circ(\delta_c),\mu_c^\circ(\delta_c))$ denote an optimal solution to $\Dc_c^{\text{lk}}(\delta_c)$. We can also solve $\Pc_c^{\text{w}}(\theta_c)$ through its dual problem since its strong duality always holds. Define $\Vc_c^{\text{w}}(\theta_c)\triangleq\{\widetilde{\Vbf}_c^{\text{w}\circ}(\theta_c)\}$
and $\Vc_c^{\text{lk}}(\delta_c)\triangleq\{\widetilde{\Vbf}_c^{\text{lk}\circ}(\delta_c)\}$ as the sets of all optimal solutions to $\Pc_c^{\text{w}}(\theta_c)$ and $\Pc_c^{\text{lk}}(\delta_c)$, respectively. By comparing the dual problems of $\Pc_c^{\text{lk}}(\delta_c)$ and $\Pc_c^{\text{w}}(\theta_c)$, the following lemma shows that, for any $\delta_c>\delta_c^{\text{w}}$, there exists $\theta_c\in[0,1)$, such that the two problems $\Pc_c^{\text{lk}}(\delta_c)$ and $\Pc_c^{\text{w}}(\theta_c)$ are equivalent.

\begin{lemma}\label{LM:deltac}
For any $\delta_c>\delta_c^{\text{w}}$, if $\theta_c=\frac{\nu_c^\circ(\delta_c)}{1+\nu_c^\circ(\delta_c)}$,
then $\Pc^{\text{w}}_c(\theta_c)$ and $\Pc_c^{\text{lk}}(\delta_c)$ are equivalent, \ie
\begin{align}
        \Vc_c^{\text{w}}\left(\frac{\nu_c^\circ(\delta_c)}{1+\nu_c^\circ(\delta_c)}\right)=\Vc_c^{\text{lk}}(\delta_c),\quad\forall\delta_c>\delta_c^{\text{w}}.\label{EQ:deltac}
\end{align}
\end{lemma}
\textit{Proof:} See Appendix \ref{APP:deltac}.

Based on Lemmas \ref{LM:fea}-\ref{LM:deltac}, we conclude in the following theorem that any optimal solution to $\Pc_c^{\text{lk}}(\delta_c)$ is also optimal for $\Pc_c^{\text{w}}(\theta_c)$ for some $\theta_c\in[0,1]$.

\begin{theorem}\label{Thm:LK}

For $\Pc_c^{\text{lk}}(\delta_c)$ being feasible, the following relations hold for $\Pc_c^{\text{w}}(\theta_c)$ and $\Pc_c^{\text{lk}}(\delta_c)$:

\begin{enumerate}
\item[\romannum{1})] For any $\delta_c>\delta_c^{\text{w}}$, there exists $\theta_c\in[0,1)$, such that $\Vc_c^{\text{lk}}(\delta_c)=\Vc_c^{\text{w}}(\theta_c)$.
\item[\romannum{2})] $\Vc_c^{\text{lk}}(\delta_c^{\text{w}})\subseteq\Vc_c^{\text{w}}(1)$.
\end{enumerate}
\end{theorem}
\textit{Proof:} See Appendix \ref{APP:LK}.

Theorem \ref{Thm:LK} shows that \textit{all} the feasible precoding solutions to $\Pc_c^{\text{lk}}(\delta_c)$ can be obtained by solving $\Pc_c^{\text{w}}(\theta_c)$, for some $\theta_c$, instead for each cell $c$. This conclusion is important since, as we will show next, $\Pc^{\text{w}}(\thetabf)$ has a semi-closed-form solution. In contrast, directly solving $\Pc^{\text{lk}}(\deltabf)$ is more complicated and does not yield such a simple semi-closed-form solution. Also, for $\Pc^{\text{lk}}(\deltabf)$, the relationship between $\deltabf$ and the system performance, \eg the average per-user rate, can be highly complicated, which adds difficulty in choosing $\deltabf$ to maximize the system performance. The selection of $\thetabf$ in $\Pc^{\text{w}}(\thetabf)$ is much easier, as we will show in our simulation study. Furthermore, Theorem \ref{Thm:LK} indicates that, for any system performance measure, the best performance achieved by solving $\Pc^{\text{w}}(\thetabf)$ is no worse than the one obtained from solving $\Pc^{\text{lk}}(\deltabf)$. As a result, for multi-cell virtualization, we can focus on $\Pc^{\text{w}}(\thetabf)$.

The above analysis can be similarly extended to the relation between $\Pc^{\text{d}}(\etabf)$ and $\Pc^{\text{w}}(\thetabf)$ with some care of technical details and hence is omitted.

\subsection{Fully Distributed Semi-Closed-Form Solution}
\label{Fully Distributed Semi-Closed-Form Solution}

We now consider solving $\Pc^\text{w}(\thetabf)$. Note that, for a practically sound virtualization design, both signal leakage and precoding deviation need to be jointly considered. Thus, in solving $\Pc^\text{w}(\thetabf)$ below, we only focus on $\theta_c\in(0,1),\forall{c}\in\Cc$. From the above discussion, we can solve $\Pc^{\text{w}}(\thetabf)$ distributively by its local precoding optimization problem $\Pc_c^{\text{w}}(\theta_c)$ at each cell $c$, without any CSI exchange across cells or central update on the transmit power from each cell that is required by conventional coordinated precoding schemes \cite{YuCOOD}\nocite{XWCOOD}-\cite{TonyCOOD}.

In Section \ref{Single-Cell MIMO WNV}, we have shown that the precoding deviation minimization problem $\Pc$ for the single-cell case has a semi-closed-form precoding solution as shown in (\ref{EQ:SC_OPT1})-(\ref{EQ:SC_OPT3}) and Proposition \ref{LM:Bsec}. In the following, we show that the weighted sum cost minimization problem $\Pc_c^{\text{w}}(\theta_c)$ for the multi-cell case can be transformed into a similar format as $\Pc$.

We first observe that the objective of $\Pc_c^{\text{w}}(\theta_c)$ can be rewritten as follows:
\begin{align}
        &(1-\theta_c)f_c(\widetilde{\Vbf}_c)+\theta_c\rho_c(\widetilde{\Vbf}_c)\notag\\
        &=(1-\theta_c)\sum_{l\neq{c},l\in\Cc}\Vert\bar{\Hbf}_{lc}\widetilde{\Vbf}_c\Vert_F^2+\theta_c\Vert\bar{\Hbf}_{cc}\widetilde{\Vbf}_c-\sqrt{P_c^{\text{w}}}\Dbf_c\Vert_F^2\notag\\
        &=\theta_c\Vert\Hbf_c^{\text{\begin{tiny}eff\end{tiny}}}\widetilde{\Vbf}_c-\sqrt{P_c^{\text{w}}}\widetilde{\Dbf}_c\Vert_F^2\label{EQ:MC_eff}
\end{align}
where we define the effective channel matrix as $\Hbf_c^{\text{\begin{tiny}eff\end{tiny}}}\triangleq\left[\beta_1\bar{\Hbf}_{1c}^H,\dots,\beta_C\bar{\Hbf}_{Cc}^H\right]^H$,
where $\beta_c=1$, $\beta_{l}=\sqrt{\frac{1-\theta_c}{\theta_c}},\forall{l}\neq{c},{l}\in\mathcal{C}$, and $\widetilde{\Dbf}_c\triangleq[\0bf,\dots,\Dbf_c^H,\dots,\0bf]^H\in\mathbb{C}^{K\times{K}_c}$. Thus, $\Pc_c^{\text{w}}(\theta_c)$ is equivalently transformed to the following problem:
\begin{align*}
        \widetilde{\Pc}_c^{\text{w}}(\theta_c):\quad\min_{\widetilde{\Vbf}_c}\quad&\Vert\Hbf_c^{\text{\begin{tiny}eff\end{tiny}}}\widetilde{\Vbf}_c-\sqrt{P_c^{\text{w}}}\widetilde{\Dbf}_c\Vert_F^2\notag\\
        \text{s.t.}~~~&\Vert\widetilde{\Vbf}_c\Vert_F^2-P_c^{\text{max}}\le0
\end{align*}
which has the same form as $\Pc$ in the single-cell case. Therefore, $\Pc_c^{\text{w}}(\theta_c)$ can be viewed as an \textit{effective} precoding deviation minimization problem in the network similar to $\Pc$. In particular, weight factor $\theta_c$ controls the significance of interfering channel $\bar{\Hbf}_{lc}$ in the effective channel $\Hbf_c^\text{\begin{tiny}eff\end{tiny}}$, leading to the trade-off between signal leakage and precoding deviation.

As a result, the optimal solution $\widetilde{\Vbf}_c^{\text{w}\circ}(\theta_c)$
to $\Pc_c^{\text{w}}(\theta_c)$ is in a semi-closed form similar to that for $\Pc$ in Section~\ref{SEC:Semi-Closed-Form Precoding Solution}, given as follows:
\begin{align}
        \widetilde{\Vbf}_c^{\text{w}\circ}(\theta_c)=\sqrt{P_c^{\text{w}}}{\Hbf_c^{\text{\begin{tiny}eff\end{tiny}}}}^\dagger\widetilde{\Dbf}_c,\label{EQ:MC_OPT}
\end{align}
if $P_c^{\text{w}}\Vert{\Hbf_c^{\text{\begin{tiny}eff\end{tiny}}}}^\dagger\widetilde{\Dbf}_c\Vert_F^2\le{P}_c^{\text{max}}$. Otherwise,
\begin{align}
        \widetilde{\Vbf}_c^{\text{w}\circ}(\theta_c)=\sqrt{P_c^\text{w}}({\Hbf_c^{\text{\begin{tiny}eff\end{tiny}}}}^H\Hbf_c^{\text{\begin{tiny}eff\end{tiny}}}+\lambda_c^\circ\Ibf)^{-1}{\Hbf_c^{\text{\begin{tiny}eff\end{tiny}}}}^H\widetilde{\Dbf}_c\label{EQ:MC_OPT2}
\end{align}
where $\lambda_c^\circ>0$ is set such that power constraint (\ref{EQ:MC_gc}) is met with equality. The search range for $\lambda_c^\circ$ is given in the following proposition. The proof is similar to the proof of Proposition \ref{LM:Bsec} and hence is omitted.

\begin{proposition}\label{LM:MC_Bsec} 
For $\widetilde{\Vbf}_c^{\text{w}\circ}(\theta_c)$ in (\ref{EQ:MC_OPT2}), the optimal Lagrange multiplier $\lambda_c^\circ$ lies in the interval $\lambda_c^\circ\in\left(0,\Vert\Hbf_c^{\text{\begin{tiny}eff\end{tiny}}}\Vert_F^2\sqrt{\frac{N_cP_c^{\text{w}}}{P_c^{\text{max}}}}\right]$.
\end{proposition}

For this distributed solution, the computational complexity for solving the subproblem $\Pc_c^{\text{w}}(\theta_c)$ is in the order of $\mathcal{O}(\min(N_c,K)^3)$, and overall is $\mathcal{O}(\sum_{c\in\Cc}\min(N_c,K)^3)$ for the original problem $\Pc^{\text{w}}(\thetabf)$ in the worst case. It is significantly less than $\mathcal{O}(\min(N,K)^3)$ for solving $\Pc^{\text{w}}(\thetabf)$ directly, \eg using an interior-point method, especially when $N_c$ and $K_c$ are large.  

So far we have obtained the precoding solution to $\Pc^{\text{w}}(\thetabf)$
for given $\thetabf$. What remains at the InP is to determine weight $\thetabf$
for the virtualization design. Note that the weighted sum cost minimization objective in $\Pc^{\text{w}}(\thetabf)$ is tailored for WNV, and therefore is not directly related to the conventional system performance metrics in terms of the data rates for non-virtualized networks. In a virtualized network, each SP has its own performance metric in generating its virtualization demand, \eg one SP could be interested in the sum rate while another SP may be more concerned about the minimum rate guarantee, for their respective sets of subscribing users. These performance targets are oblivious to the InP, who is only concerned about meeting the virtualization demands provided by the SPs. Compared with conventional non-virtualized networks, our proposed virtualized precoding solution caters to different service needs of SPs, allowing the network to be shared in a more flexible manner. 

Even in the case all SPs use a common performance metric and the InP uses it (\eg sum rate or minimum rate) to optimize the weight $\thetabf$, the problem is still very challenging, as the objective may be highly non-convex w.r.t. $\thetabf$. Let $\ybf_c=\sum_{l\in\Cc}\ybf_{cl}$ be the actual received signal at the $K_c$ users in cell $c$. The  global received signal at all $K$ users in the network  $\ybf=[\ybf_1^H,\dots\ybf_C^H]^H$ in compact form is given by
\begin{align}
        \ybf=\Hbf\Vbf\xbf\label{EQ:y}
\end{align}
where $\Hbf=[\widetilde{\Hbf}_1,\dots,\widetilde{\Hbf}_C]$, $\Vbf=\blkdiag\{\widetilde{\Vbf}_1,\dots,\widetilde{\Vbf}_C\}$, and $\xbf=[\xbf_1^H,\dots,\xbf_C^H]^H$ with  $\mathbb{E}\{\xbf\xbf^H\}=\Ibf$.
The virtual received signal at all $K$ users in the network $\widetilde{\ybf}=[\widetilde{\ybf}_{1}^H,\dots,\widetilde{\ybf}_{C}^H]^H$ is given by
\begin{align}
        \widetilde{\ybf}=\mathbf{P}^{\text{w}}\Dbf\xbf\label{EQ:tildey}
\end{align}
where $\mathbf{P}^{\text{w}}=\blkdiag\{\sqrt{P_1^{\text{w}}}\Ibf,\dots,\sqrt{P_C^{\text{w}}}\Ibf\}$ and  $\Dbf=\blkdiag\{\Dbf_1,\dots,\Dbf_C\}$. Note that the virtual received signal $\widetilde{\ybf}$ does not consider either the intra-cell inter-SP interference or the inter-cell interference. The expected deviation of received signals at all $K$ users, between the actual precoding by the InP and precoding demand by the SPs, is given by
\begin{align*}
        &\mathbb{E}_{\xbf}\{\Vert\ybf-\widetilde{\ybf}\Vert_F^2\}=\sum_{c\in\Cc}\Vert\widetilde{\Hbf}_c\widetilde{\Vbf}_c-\sqrt{P_c^{\text{w}}}\widetilde{\Dbf}_c\Vert_F^2\notag\\
        &=\sum_{c\in\Cc}\left(\sum_{l\neq{c},l\in\Cc}\Vert\bar{\Hbf}_{lc}\widetilde{\Vbf}_c\Vert_F^2+\Vert\bar{\Hbf}_{cc}\widetilde{\Vbf}_c-\sqrt{P_c^{\text{w}}}\Dbf_c\Vert_F^2\right)\notag\\
        &=\sum_{c\in\Cc}\left(f_c(\widetilde{\Vbf}_c)+\rho_c(\widetilde{\Vbf}_c)\right)
\end{align*}
which is exactly the objective of $\Pc^{\text{w}}(\mathbf{\frac{1}{2}})$, where $\mathbf{\frac{1}{2}}=[\frac{1}{2},\dots,\frac{1}{2}]$. Therefore, at $\thetabf=\mathbf{\frac{1}{2}}$, the InP is equivalently minimizing the \textit{global} total precoding deviation. When it is zero, the precoding demands of all SPs in the network are met, without either the intra-cell inter-SP interference or the inter-cell interference. This suggests that $\thetabf=\mathbf{\frac{1}{2}}$ is a special weight vector from the InP's perspective of whole network operation. Indeed, in Section \ref{Performance Evaluation}, we numerically show that $\thetabf=\mathbf{\frac{1}{2}}$ can achieve close to optimal system performance for the two metrics of sum rate and minimum rate. However, since the global total precoding deviation does not necessarily indicate the individual cell performance, we emphasize that  $\thetabf$ may be designed to control the relative performance across cells.  

\subsection{Fully Distributed Virtual Transmit Power Allocation $P_c^{\normalfont\text{w}}$}

\setcounter{equation}{30}
\begin{figure*}[b]
        \hrule
        \begin{align}
                \text{SINR}_{cmk}=\frac{|[\Hbf_{cc}^m\Vbf_c^m]_{k,k}|^2}{\displaystyle\sum_{i\neq{k},i\in\Kc_{c}^m}|[\Hbf_{cc}^m\Vbf_c^m]_{k,i}|^2
                +\sum_{j\neq{m},j\in\Mc}\sum_{i\in\Kc_c^j}|[\Hbf_{cc}^m\Vbf_c^j]_{k,i}|^2
                +\sum_{l\neq{c},l\in\Cc}\sum_{j\in\Mc}\sum_{i\in\Kc_l^j}|[\Hbf_{cl}^m\Vbf_{l}^j]_{k,i}|^2+\sigma_{n}^2}.\label{EQ:SINR}
        \end{align}
\end{figure*}

The solution obtained for $\widetilde{\Pc}_c^\text{w}(\theta_c)$ so far is for given virtual transmit power $P_c^{\text{w}}\le{P}_c^\text{max}$, for ${c}\in\Cc$. Recall from the discussion in Section~\ref{SEC:Virtual Transmit Power Allocation} that $P_c^{\text{w}}$ is intended to regulate interference suppression and each SP's virtualization demand maximization. It needs to be properly determined to reflect the actual desired service quality, but is challenging to be optimized. Instead of a computationally expensive exhaustive search for each optimal $P_c^{\text{w}}$, we extend the virtual power allocation scheme proposed in Section~\ref{SEC:Virtual
Transmit Power Allocation} to the multi-cell case. We propose an intuitive and computationally efficient virtual transmit power allocation scheme for $\{P_c^{\text{w}}\}$ for the virtualized multi-cell MIMO system. Since $\widetilde{\Pc}^\text{w}_c(\theta_c)$ and $\Pc$ have the same format, similar to (\ref{EQ:SC_0Dev}) and (\ref{EQ:SC_Pw}) in the single-cell case, the maximum value of $P_c^{\text{w}}$ for $\widetilde{\Vbf}_c^{\circ}(\theta_c)$ in (\ref{EQ:MC_OPT}) to satisfy $(\ref{EQ:MC_gc})$ is given by
\begin{align}
        P_c^{\text{w}\circ}=\min\left\{\frac{P_c^{\text{max}}}{\Vert{\Hbf_c^{\text{\begin{tiny}eff\end{tiny}}}}^\dagger\widetilde{\Dbf}_c\Vert_F^2},P_c^{\text{max}}\right\},\quad{c}\in\mathcal{C}.\label{EQ:MC_Pw}
\end{align}
Note that, given fixed $\widetilde{\Vbf}_l$ for all $l\neq{c}$, with $\widetilde{\Vbf}_c^{\text{w}\circ}(\theta_c)$ in (\ref{EQ:MC_OPT}), the SINR of each user in cell $c$ is monotonically increasing with the virtual transit power $P_c^{\text{w}}$. As such,  $P_c^{\text{w}\circ}$ in (\ref{EQ:MC_Pw}) greedily maximizes the SINRs of the $K_c$ users in cell $c$. Therefore, we propose to use $P_c^{\text{w}\circ}$ in (\ref{EQ:MC_Pw}) to the solution to $\widetilde{\Pc}^\text{w}_c(\theta_c)$ as  $\widetilde{\Vbf}_c^{\text{w}\circ}(\theta_c)=\sqrt{P_c^{\text{w}\circ}}{\Hbf_c^{\text{\begin{tiny}eff\end{tiny}}}}^\dagger\widetilde{\Dbf}$.

Note that $P_c^{\text{w}\circ}$ depends, through $\Hbf_c^\text{\begin{tiny}eff\end{tiny}}$, on $\theta_c$ and all the channels to users in cell $c$. Also note that if $N_c\ge K$, \ie there are sufficient degrees of freedom, the objective function in $\widetilde{\Pc}_c^{\text{w}}(\theta_c)$ is under-determined and the optimal value is zero, \ie the InP can achieve zero leakage and inter-SP interference using the precoding $\widetilde{\Vbf}_c^{\text{w}\circ}(\theta_c)$ in (\ref{EQ:MC_OPT}) and achieves complete service isolation desired for WNV. In this case, the proposed virtual transmit power allocation $P_c^{\text{w}\circ}$ in (\ref{EQ:MC_Pw}) at cell $c$ is the maximum power to meet the SP precoding demand while nulling the inter-SP interference in cell $c$ without generating any signal leakage to the other cells. If $N_c<K$, there is not enough degrees of freedom for the InP to eliminate signal leakage and inter-SP interference at the same time.

Similar to the proposed virtual transmit power $P^{\text{w}\circ}$ in (\ref{EQ:SC_Pw}) for the single-cell case, the choice of $P_c^{\text{w}\circ}$ in (\ref{EQ:MC_Pw}) leads to a closed-form precoding solution $\widetilde{\Vbf}_c^{\text{w}\circ}(\theta_c)$ given in (\ref{EQ:MC_OPT}), instead of (\ref{EQ:MC_OPT2}), which is for $P_c^{\text{w}}>P_c^{\text{w}\circ}$. Through simulation,  we will show that this choice of $P_c^{\text{w}\circ}$ (and $\widetilde{\Vbf}_c^{\text{w}\circ}(\theta_c)$) gives close to optimal system performance in terms of average rate or minimum rate, among all possible values of $P_c^{\text{w}}$. In particular, setting $P_c^{\text{w}} > P_c^{\text{w}\circ}$ may lead to much degraded system performance.


\section{Simulation Results}
\label{Performance Evaluation}

Our coordinated MIMO virtualized precoding design trades-off the signal leakage
and precoding deviation at the SPs to reach certain desired system performance. In our simulation, we consider two important system performance measures commonly used for non-virtualized networks, the sum rate and the minimum user rate. Our first system performance metric is the average per-user rate in the network defined as
\setcounter{equation}{29}
\begin{align}
        \bar{R}(\Vbf)\triangleq\frac{1}{K}\sum_{c\in\Cc}\sum_{m\in\Mc}\sum_{k\in\Kc_c^m}\log_2(1+\text{SINR}_{cmk})\label{EQ:Rbar}
\end{align}
where $\Kc_c^m=\{1,\dots,K_c^m\}$, and $\text{SINR}_{cmk}$ is the SINR of the $k$-th user of SP $m$ in cell $c$ given by (\ref{EQ:SINR}). The second performance metric is the averaged minimum rates of all SPs, given by
\setcounter{equation}{31}
\begin{align}
        \bar{R}_{\text{min}}(\Vbf)\triangleq\frac{1}{CM}\sum_{c\in\Cc}\sum_{m\in\Mc}\min_{k\in\Kc_c^m}\log_2(1+\text{SINR}_{cmk}).\label{EQ:barRmin}
\end{align} 
It is the minimum rate at each virtual cell of an SP, averaged over all SPs and all cells, and normalized by the system bandwidth bandwidth. Both $\bar{R}(\Vbf)$ and $\bar{R}_{\text{min}}(\Vbf)$ are highly non-convex w.r.t. precoding matrix $\Vbf$, and thus are challenging to optimize, even in non-virtualized networks.

\subsection{Simulation Setup}

\begin{figure}[!t]
\centering
\vspace{0mm}
\subfloat[$N_c=32<K=56$.]
{\includegraphics[width=1\linewidth,trim=22 0 22 0,clip]{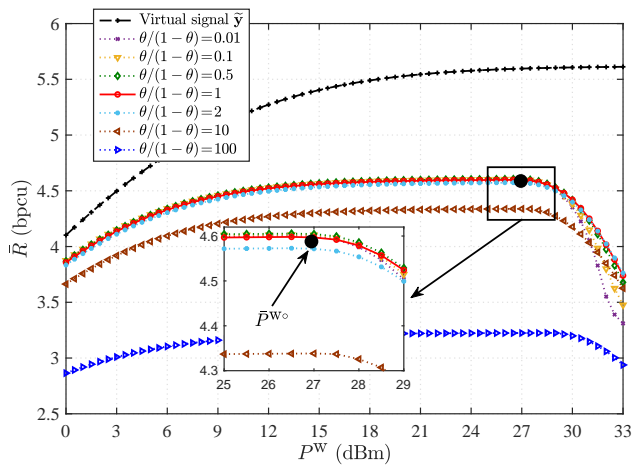}}
\vspace{0mm}
\subfloat[$N_c=64>K=56$.]
{\includegraphics[width=1\linewidth,trim=22 0 22 0,clip]{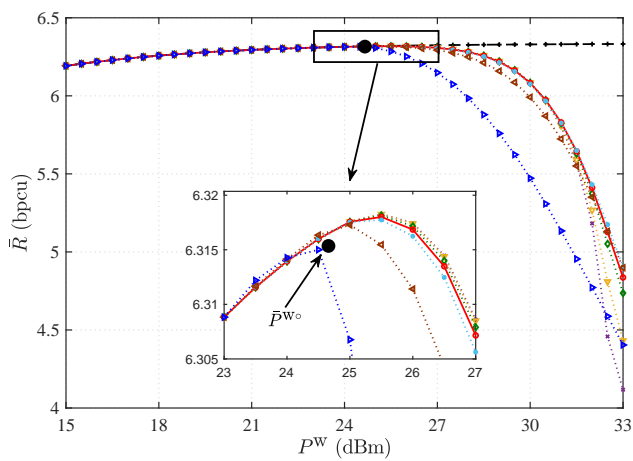}}
\caption{$\bar{R}$ vs. $P^{\text{w}}$ and $\theta$ when all SPs adopt
MRT precoding (the same legend in Fig.~2(a) also applies to Fig.~2(b)).}\label{Fig:MRT}
\vspace{0mm}
\end{figure}

\begin{figure}[!t]
\centering
\vspace{0mm}
\subfloat[$N_c=32<K=56$.]
{\includegraphics[width=1\linewidth,trim=22 0 22 0,clip]{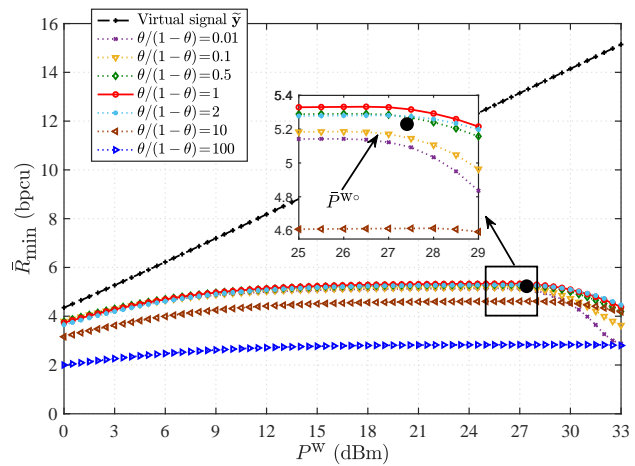}}
\vspace{0mm}
\subfloat[$N_c=64>K=56$.]
{\includegraphics[width=1\linewidth,trim=22 0 22 0,clip]{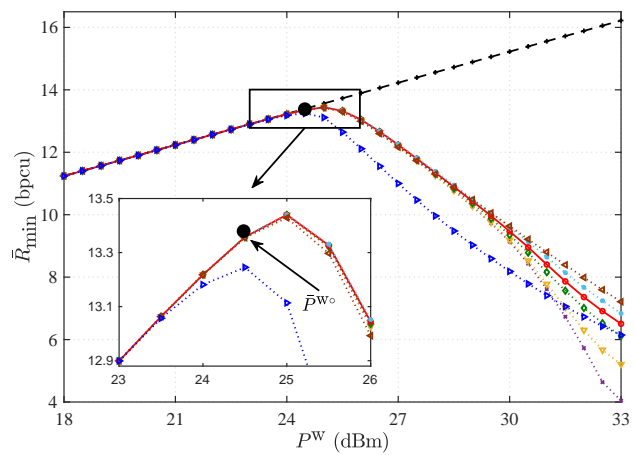}}
\caption{$\bar{R}_{\text{min}}$ vs. $P^{\text{w}}$ and $\theta$ when all SPs adopt ZF precoding  (the same legend in Fig.~3(a) also applies to Fig.~3(b)).}\label{Fig:ZF}
\vspace{0mm}
\end{figure}

We consider that an InP owns a MIMO cellular network consisting of $C=7$ urban hexagon micro cells. Each cell $c$ has radius $R_c=500$~m.
The InP serves $M=4$ SPs, and each SP $m$ serves $K_c^m=2$ users in cell $c$. Following the standard LTE specifications \cite{LTEP}, we set the following default parameters. The maximum transmit power to $P_c^{\text{max}}=33$ dBm, noise spectral density $N_0=-174$~dBm/Hz, and noise figure $N_F=10$~dB.  We focus on transmission over bandwidth $B_W=15$ kHz. The channel between BS $c$ and user $k$ is modeled as $\hbf_c^k=\sqrt{\beta_c^k}\gbf_c^k,\forall{c}\in\Cc,\forall{k}\in\Kc$, where $\Kc=\{1,\dots,K\}$, $\gbf_c^k\sim\mathcal{CN}(\0bf,\Ibf)$, $\beta_c^k [\text{dB}]=-31.54-33\log_{10}(d_c^k)-\psi_c^k$ represents the path-loss and shadowing with $d_c^k$ being the distance in kilometers from the BS in cell $c$ to user $k$ and $\psi_c^k\sim\mathcal{CN}(0,\sigma_\psi^2)$ being the shadowing with $\sigma_\psi=8$ dB. To study the impact of inaccurate
CSI, for channel state~$\hbf_c^k$, we generate its CSI error through $\mathcal{CN}(\0bf,e_{\mathbf{H}}^2\beta_c^k\Ibf)$. 

For our performance study, we consider that each SP $m$ adopts either MRT or ZF precoding, two commonly used precoding schemes in MIMO systems, to design its normalized virtual precoding matrix. They are given by
\begin{align}
\mathbf{W}_c^m=
\left\{\begin{matrix}
        \varpi_{c}^m{\Hbf_{cc}^m}^H, &\text{for~MRT}\\
        \varpi_{c}^m{\Hbf_{cc}^m}^H(\Hbf_{cc}^m{\mathbf{H}_{cc}^m}^H)^{-1},&\text{for~ZF}
\end{matrix}\right.\label{eq:Wccm}
\end{align}
where $\varpi_{c}^m$ is a power normalization factor such that $\Vert\Wbf_c^m\Vert_F^2=1$.
We assume that the InP allocates equal virtual transmit power to the SPs in each cell for fair resource allocation among the SPs, \ie $\alpha_c^m=\frac{1}{M},\forall{m}\in\Mc,\forall{c}\in\Cc$.
Note that the optimal max-min SINR precoder under the single-cell setting is in fact a MMSE precoder \cite{SCSINR}. In the high signal-to-noise ratio (SNR) region, ZF precoding approaches  the MMSE precoder  \cite{highSNR}. Indeed, in our simulation, we observe negligible performance difference between the case when all SPs adopt the max-min SINR precoding \cite{SCSINR} and the case when all SPs adopt ZF precoding. Therefore, in the following, if user fairness is of interest at an SP, we assume it adopts ZF precoding to design its virtualization service demand.

\subsection{Impact of Virtual Transmit Power $P_c^{\normalfont\text{w}}$}
\label{Sec:Impact of Virtual Transmit Power}

The search space of $\{P_c^{\text{w}},\theta_c\}$ in $\Pc^{\text{w}}(\thetabf)$ to reach the optimal $\bar{R}(\Vbf)$ or $\bar{R}_{\text{min}}(\Vbf)$ is very large. For the purpose of illustration, we assume that the InP sets the same virtual transmit power $P_c^{\text{w}}=P^{\text{w}}$ and the same weight factor $\theta_c=\theta$, for all $c\in\mathcal{C}$. We vary $P^{\text{w}}$ and $\theta$ to study the impacts of signal leakage and precoding deviation on the performance of our algorithm, where we obtain $\bar{R}(\mathbf{V})$ or $\bar{R}_{\text{min}}(\mathbf{V})$ for given $P^\text{w}$ and $\theta$ based on the solution $\{\widetilde{\Vbf}_c^{\text{w}\circ}(\theta)\}$
in (\ref{EQ:MC_OPT}) or (\ref{EQ:MC_OPT2}).  Fig.~\ref{Fig:MRT} shows the average per-user rate $\bar{R}$ versus $P^{\text{w}}$ for different values of $\theta$, when all SPs adopt MRT precoding. We show the \textit{virtual received signal} $\widetilde{\ybf}$ in (\ref{EQ:tildey}) as the desired signal by the SPs based on their virtual precoding $\{\Wbf_c^m\}$ in~(\ref{eq:Wccm}).

Fig.~2(a) shows the case when the number of antennas $N_c =32$ and the total users in the network $K=56$, where there is not enough degrees of freedom to eliminate signal leakage and achieve zero precoding deviation at the same time.  Note that $\widetilde{\mathbf{y}}$ in (\ref{EQ:tildey}) is only what the SP wants, without considering either the inter-SP interference or the inter-cell interference. Therefore, $\bar{R}$ achieved by $\widetilde{\mathbf{y}}$ is higher than that achieved by the \textit{actual} received signal $\mathbf{y}$ in (\ref{EQ:y}) due to the proposed precoding scheme that considers both the inter-SP and inter-cell interference. When $N_c<K$, we observe that $\bar{R}$ does not scale with the virtual transmit power $P^{\text{w}}$. This is because there are not enough degrees of freedom in the system for the InP to mitigate the interference while satisfying the SPs' demands. The system becomes interference-limited. As a result, increasing power $P^{\text{w}}$ (increasing virtualization demand) does not help to improve the system performance. As the virtual transmit power $P^{\text{w}}$ increases, we observe that $\bar{R}$ increases first and then decreases at a much faster rate. This implies that allocating the maximum transmit power as the virtual power to the SPs at each cell $c$, \ie $P_c^{\text{w}}=P_c^{\text{max}},\forall{c}\in\mathcal{C}$ (as used in \cite{Mpaper}, \cite{GLOBECOM19}\nocite{INFOCOM20}-\cite{SPAWC20}), leaves limited freedom to the InP for interference suppression in the precoding design, and this in turn may lead to severe system performance degradation.

Fig.~2(b)  shows the opposite case when $N_c=64$ and $K =56$. As $N_c > K$, with sufficient degrees of freedom, the system performance gap to the one under the virtual signal $\widetilde{\mathbf{y}}$ is drastically reduced (compared with Fig.~2(a)). As the InP-designed precoding can eliminate the signal leakage to other cells and null the inter-SP interference, when the virtual transmit power $P_c^{\text{w}}$ is low, the actual received signal $\mathbf{y}$ is identical to the desired virtual signal $\widetilde{\mathbf{y}}$, leading to identical $\bar{R}$. Furthermore, we observe that setting the weight factor $\theta=\frac{1}{2}$ yields $\bar{R}$ that is close to the maximum among different values of $\theta$. 

For both Fig.~2(a) and Fig.~2(b), we indicate the performance $\bar{R}$ at $\bar{P}^{\text{w}\circ}=\frac{1}{C}\sum_{c\in\mathcal{C}}P_c^{\text{w}\circ}$
as the averaged value of the proposed virtual transmit power $\{P_c^{\text{w}\circ}\}$
in (\ref{EQ:MC_Pw}). It is interesting to observe that, in both plots, $\bar{R}$ achieved by $\{P_c^{\text{w}\circ}\}$ is close to the maximum value of $\bar{R}$.

Fig.~\ref{Fig:ZF} shows the averaged minimum rates $\bar{R}_{\text{min}}$ of all SPs versus $P^{\text{w}}$ for different values of $\theta$, when all SPs adopt ZF precoding. Compared with MRT precoding, $\bar{R}_{\text{min}}$ achieved by ZF precoding is much higher. This is because our system is operated at high SNR, and ZF precoding is close to optimal precoding in this region. Similar to Fig 2(a), in Fig. 3(a), when $N_c<K$, there are insufficient degrees of freedom to mitigate interference, and the system is interference-limited. Thus, $\bar{R}_{\text{min}}$ does not scale with $P^{\text{w}}$. Similar to the MRT precoding case, setting $\theta=\frac{1}{2}$ yields close to the maximum value of $\bar{R}_{\text{min}}$ among different values of $\theta$, and $\bar{R}_{\text{min}}$ achieved with $\{P_c^{\text{w}\circ}\}$ is close to the maximum value of $\bar{R}_{\text{min}}$. 

We have shown that setting weight factor $\theta=\frac{1}{2}$ is close-to-optimal and the proposed virtual transmit power $\{P_c^{\text{w}\circ}\}$ is effective, for both the average per-user rate $\bar{R}$ and the averaged minimum rates of the SPs $\bar{R}_{\text{min}}$. As such, in practice, the InP can simply set $\theta_c=\frac{1}{2}$ and allocate $P_c^{\text{w}\circ}$ in (\ref{EQ:MC_Pw}) to each cell $c\in\mathcal{C}$. In this case, the weighted sum minimization problem $\Pc^{\text{w}}(\mathbf{\frac{1}{2}})$ has a closed-form solution $\{\widetilde{\Vbf}_c^{\text{w}\circ}(\frac{1}{2})\}$ in (\ref{EQ:MC_OPT}).

\subsection{Benefit of Service Isolation via Spatial Virtualization}

Based on the results above, in the following simulation, we use the closed-form precoding solution $\widetilde{\Vbf}_c^{\text{w}\circ}(\frac{1}{2})$ in (\ref{EQ:MC_OPT}) with the proposed virtual transmit power $P_c^{\text{w}\circ}$ in (\ref{EQ:MC_Pw}), for each cell $c\in\mathcal{C}$. We assume each SP $m$ adopts ZF precoding to design its normalized virtual precoding $\mathbf{W}_c^m$ in each cell $c\in\mathcal{C}$ and focus on the study of the averaged minimum rates of all SPs $\bar{R}_{\text{min}}$.

For a performance upper bound, we  consider the idealized \textit{cooperative} precoding, which is highly complicated to implement but can substantially outperform the more practical coordinate precoding approach that we consider in this work. Furthermore, since there is no low-complexity solution to cooperative precoding with per-cell transmit power constraints, we resort to assuming cooperative precoding with a sum power constraint over all cells, which further favors its performance. In the case of ZF precoding, this is given by
\begin{align}
        \Vbf_{\text{ZF}}=\varpi\Hbf^H(\Hbf\Hbf^H)^{-1}\label{EQ:CompZF}
\end{align}
where $\varpi$ is a power normalization factor such that $\Vert\Vbf_{\text{ZF}}\Vert_F^2=\sum_{c\in\Cc}P_c^{\text{max}}$. Note that $\Vbf_{\text{ZF}}$ in (\ref{EQ:CompZF}) requires sharing both the global channel state $\mathbf{H}$ and the global transmit signal $\mathbf{x}$ at each cell $c\in\mathcal{C}$, while the proposed coordinated precoding uses only the local channel state $\widetilde{\mathbf{H}}_c$ and the local transmit signal $\mathbf{x}_c$. Since the system is operated at high SNR, $\Vbf_{\text{ZF}}$ in (\ref{EQ:CompZF}) is close to optimal precoding.

We also consider service isolation via orthogonal bandwidth allocation, which is commonly adopted in existing literature \cite{rp}\nocite{EE17}-\cite{V5G}, \cite{WNVNOMA}. Specifically, we consider a  frequency division (FD) scheme that allocates equal bandwidth $\frac{B_W}{M}$ to each SP~$m$. We apply the proposed closed-form coordinated precoding solution to each SP. This is a special case of a single SP in our general solution $\{\widetilde{\Vbf}_c^{\text{w}\circ}(\frac{1}{2})\}$
in (\ref{EQ:MC_OPT}). This precoding scheme uses the local CSI to  minimize the inter-cell signal leakage while meeting each SP's demand. It can be considered as an FD leakage minimization scheme for WNV. Note that the rate for each SP is normalized by the system bandwidth $B_{W}$.

\begin{figure}[!t]
\centering
\vspace{-0mm}
\subfloat[$\bar{R}_{\text{min}}$ vs. $N_c$ with $K_c=8$ and different values
of $R_c$.\label{Fig:Comp_Nc}]
{\includegraphics[width=1\linewidth,trim=22 0 22 0,clip]{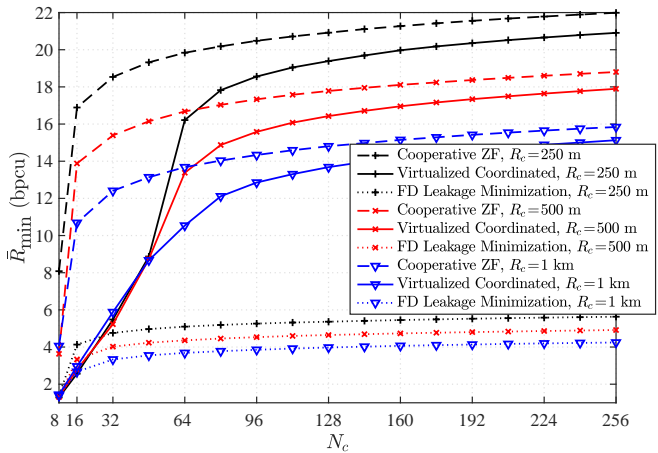}}
\vspace{0mm}
\subfloat[$\bar{R}_{\text{min}}$ vs. $K_c$ with $N_c=128$ and different values
of $R_c$.\label{Fig:Comp_Kc}]
{\includegraphics[width=1\linewidth,trim=22 0 22 0,clip]{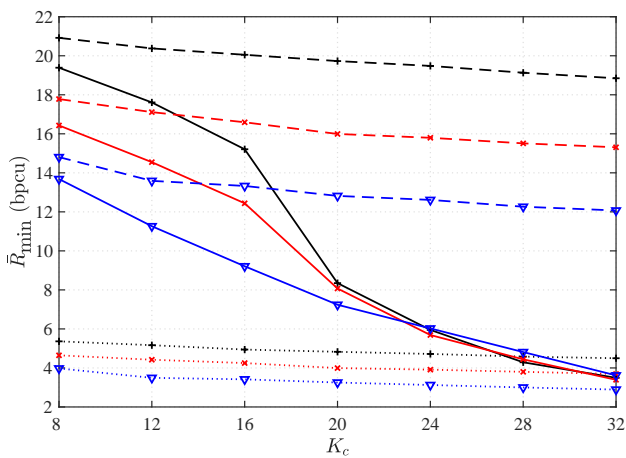}}
\caption{Comparison of $\bar{R}_{\text{min}}$ among different precoding schemes
under perfect CSI (the same legend in Fig.~4(a) also applied to Fig.~4(b)).}\label{Fig:Comp}
\vspace{0mm}
\end{figure}

Fig.~\ref{Fig:Comp} shows the performance comparison between the proposed virtualized coordinated precoding, cooperative ZF precoding, and FD leakage minimization precoding under perfect CSI. Fig.~4(a) shows the impact of $N_c$ on the system performance with fixed number of users per cell $K_c=8$. In Fig.~4(b), we examine the impact of user density by varying $K_c$ with the BS antennas fixed at $N_c=128$. When the BSs are equipped with enough antennas relative to the total users in the network, \ie $N_c\ge{K},\forall{c}\in\mathcal{C}$, as $N_c$ increases, the performance achieved by our proposed virtualized coordinated precoding grows closer to that of the idealized cooperative precoding. When the number of BS antennas is small, \ie $N_c<K$, the proposed virtualized coordinated precoding does not have enough degrees of freedom to mitigate the interference among the  SPs and cells, which leads to noticeably performance degradation. We also observe from both Fig.~4(a) and Fig.~4(b) that, as the cell size increases, the performance gap to that of the cooperative ZF precoding reduces. This is because there is less inter-cell interference for the InP to control and thus more virtual transmit power can be allocated to the SPs.

\begin{figure}[!t]
\centering
\vspace{-0mm} 
\includegraphics[width=1\linewidth,trim=22 0 22 0,clip]{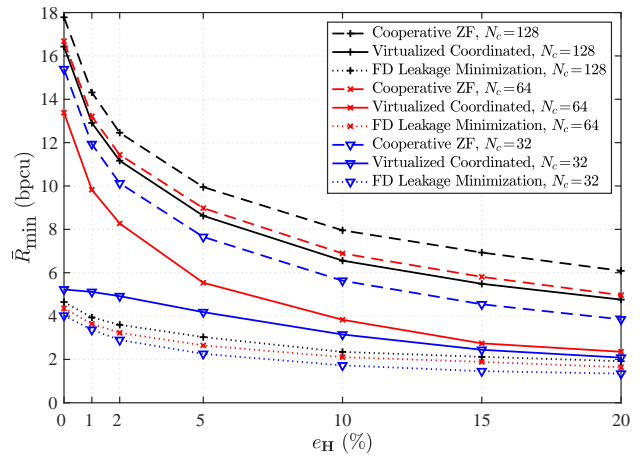}
\vspace{-6.8mm}
\caption{Comparison of $\bar{R}_{\text{min}}$ among different precoding schemes
under inaccurate CSI.}\label{Fig:Comp_eH}
\vspace{0mm}
\end{figure}

When the number of antennas is large, the proposed virtualized coordinated precoding substantially outperforms the FD leakage minimization scheme. This demonstrates the effectiveness of the proposed spatial isolation approach, with simultaneous sharing of all the frequency channel resources among SPs. Note that when there are not enough antennas for spatial isolation, \eg the case of $N_c=16<K=56$ in Fig.~4(a), the user received SINR is low due to high interference. In this regime, applying FD can be more effective than performing spatial isolation with full bandwidth, by isolating interference to increase SINR.

Fig.~\ref{Fig:Comp_eH} shows the impact of imperfect CSI on the system performance. As the number of antennas $N_c$ increases, the performance gap of our proposed virtualized coordinated precoding to that of cooperative precoding deceases, indicating the robustness of our design to CSI inaccuracy for coordinated multi-cell WNV. We observe that the proposed virtualized coordinated precoding substantially outperforms the FD signal minimization precoding in a wide range of CSI inaccuracy levels, indicating the performance gain of spatial virtualization over the spectrum isolation schemes for MIMO WNV \cite{rp}\nocite{EE17}-\cite{V5G}, \cite{WNVNOMA}.

\section{Conclusion}
\label{Conclusion}

In this paper, we considered the design of MIMO WNV to achieve service isolation among the SPs in a multi-cell scenario, where the InP decides the transmitter precoding based on SPs' individual service demands. To the best of our knowledge, this is the first work to achieve spatial virtualization in a multi-cell MIMO system with simultaneous utilization of all antennas and channel resources, while managing both the inter-SP and inter-cell interference. We show that the resultant coordinated precoding optimization problem, to minimize a weighted sum of signal leakage and precoding deviation under per-cell transmit power limits,  can be decomposed into per-cell subproblems, leading to a fully distributed semi-closed-form solution at each cell. We also propose a low-complexity virtual transmit power allocation scheme for each SP's virtual service demand to regulate between interference elimination and virtual demand maximization. Simulation results demonstrate that the system performance of a virtualized network enabled by our proposed solution is substantially higher than that of the common frequency isolation alternative,  and it can approach the performance of an idealized cooperative scheme when the number of antennas becomes large.


\appendices

\section{Proof of Proposition \ref{LM:Bsec}}\label{APP:Bsec}

\textit{Proof:} Let $\Hbf^H\Hbf+\lambda^{\circ}\Ibf=\Ubf\boldsymbol{\Sigma}\Ubf$,
where $\boldsymbol{\Sigma}=\diag\{\sigma_{1}+\lambda^{\circ},\dots,\sigma_{N}+\lambda^{\circ}\}$
with $\sigma_n$ being the eigenvalues of $\Hbf^H\Hbf$, and $\Ubf$ is a unitary
matrix. If $\lambda^{\circ}>0$, from (\ref{EQ:SC_OPT3}), we have
\begin{align}
        \Vert\Vbf^{\circ}\Vert_F^2&=P^\text{w}\Vert(\Hbf^H\Hbf+\lambda^{\circ}\Ibf)^{-1}\Hbf^H\Dbf\Vert_F^2\notag\\
        &\stackrel{(a)}{\leq}P^\text{w}\Vert(\Hbf^H\Hbf+\lambda^{\circ}\Ibf)^{-1}\Vert_F^2\Vert\Hbf\Vert_F^2\Vert\Dbf\Vert_F^2\notag\\
        &\stackrel{(b)}{\leq}P^\text{w}\Vert(\Hbf^H\Hbf+\lambda^{\circ}\Ibf)^{-1}\Vert_F^{2}\Vert\Hbf\Vert_F^{4}\notag\\
        &\stackrel{(c)}{\leq} P^\text{w}\Vert\Hbf\Vert_F^4\frac{N}{\lambda^{\circ2}}\label{EQ:lambda}
\end{align}
where $(a)$ follows from $\Vert\Abf\Bbf\Vert_F\leq\Vert\Abf\Vert_F\Vert\Bbf\Vert_F$,
$(b)$ is because $\alpha^m\leq1$ and $\Vert\mathbf{W}^m\Vert_F^2=1,\forall{m}\in\Mc$,
and thus
\begin{align*}
        \Vert\Dbf\Vert_F^2=\sum_{m\in\Mc}\alpha^m\Vert\Hbf^m\Wbf^m\Vert_F^2\leq\sum_{m\in\Mc}\Vert\Hbf^m\Vert_F^2=\Vert\Hbf\Vert_F^2,
\end{align*}
and $(c)$ is because $\sigma_{n}\geq0, n=1,\dots,N$ and thus
\begin{align*}
        \Vert(\Hbf^H\Hbf+\lambda^{\circ}\Ibf)^{-1}\Vert_F^2=\sum_{n=1}^N\frac{1}{(\sigma_{n}+\lambda^{\circ})^2}\le\frac{N}{\lambda^{\circ2}}.
\end{align*}
Since by (\ref{EQ:SC_KKT4}), the equality holds for (\ref{EQ:SC_KKT2}) at
optimality, following (\ref{EQ:lambda}), we have $\lambda^{\circ}\le\Vert\Hbf\Vert_F^2\sqrt{\frac{NP^\text{w}}{P^{\text{max}}}}$.
\endIEEEproof

\section{Proof of Lemma \ref{LM:fea}}\label{APP:fea}
\textit{Proof:} We first prove ``only if" by contradiction, \ie $\delta_c\ge\delta_c^{\text{w}}$ if $\Pc_c^{\text{lk}}(\delta_c)$ is feasible. Suppose there exists a $\delta_c'<\delta_c^{\text{w}}$ such that $\Pc^{\text{lk}}_c(\delta_c')$ is feasible. We have $\rho_c(\widetilde{\Vbf}_c^{\text{lk}\circ}(\delta_c'))\leq\delta_c'$
and $\Vert\widetilde{\Vbf}_c^{\text{lk}\circ}(\delta_c')\Vert_F^2-P_c^{\text{max}}\le0$. By (\ref{EQ:MC_gc}), $\widetilde{\Vbf}_c^{\text{lk}\circ}(\delta_c')$ is also a feasible solution to $\Pc_c^{\text{w}}(1)$. From the above assumption, we also have $\rho_c(\widetilde{\Vbf}_c^{\text{lk}\circ}(\delta_c'))<\rho_c(\widetilde{\mathbf{V}}_c^{\text{w}\circ}(1))=\delta_c^{\text{w}}$,
which contradicts the fact that $\widetilde{\mathbf{V}}_c^{\text{w}\circ}(1)$ is an optimal solution to $\Pc_c^{\text{w}}(1)$.

To prove ``if", note that, when $\delta_c\ge\delta_c^{\text{w}}$, from the definition of $\delta_c^{\text{w}}$ in  (\ref{EQ:defdeltac-}), we have $\rho_c(\widetilde{\Vbf}_c^{\text{w}\circ}(1))\le\delta_c$. Also, $\widetilde{\Vbf}_c^{\text{w}\circ}(1)$ satisfies the transmit power constraint (\ref{EQ:MC_gc}). Thus, $\widetilde{\Vbf}_c^{\text{w}\circ}(1)$ is a feasible solution to $\Pc_c^{\text{lk}}(\delta_c)$  for any $\delta_c\ge\delta_c^{\text{w}}$.
\endIEEEproof

\section{Proof of Lemma \ref{LM:dual}}\label{APP:dual}

\textit{Proof:} Since $\Pc_c^{\text{lk}}(\delta_c)$ is convex for $\delta_c>\delta_c^{\text{w}}$, we prove strong duality by showing the Slater's condition holds. We prove the lemma by considering the following
two cases.

1) $\Vert\widetilde{\Vbf}_c^{\text{lk}\circ}(\delta_c^{\text{w}})\Vert_F^2<P_c^{\text{max}}$: Since $\rho_c(\widetilde{\Vbf}_c^{\text{lk}\circ}(\delta_c^{\text{w}}))=\delta_c^{\text{w}}<\delta_c$,
$\widetilde{\Vbf}_c^{\text{lk}\circ}(\delta_c^{\text{w}})$ satisfies the Slater's condition.

2) $\Vert\widetilde{\Vbf}_c^{\text{lk}\circ}(\delta_c^{\text{w}})\Vert_F^2=P_c^{\text{max}}$: From the convexity of the power constraint function in (\ref{EQ:MC_gc}), for any $t\in(0,1]$, we have
\begin{align}
        &\Vert t\mathbf{0}+(1-t)\widetilde{\mathbf{V}}_c^{\text{lk}\circ}(\delta_c^{\text{w}})\Vert_F^2-P_c^{\text{max}}\notag\\
        &\leq t \left(\Vert\mathbf{0}\Vert_F^2-P_c^{\text{max}}\right)+(1-t)\left(\Vert\widetilde{\Vbf}_c^{\text{lk}\circ}(\delta_c^{\text{w}})\Vert_F^2-P_c^{\text{max}}\right)\notag\\
        &=-tP_c^{\text{max}}<0.\label{EQ:conv1}
\end{align}
Similarly, from the convexity of $\rho_c(\widetilde{\Vbf}_c)$ in (\ref{EQ:MC_rho}), for any $t\in(0,1]$, we have
\begin{align}
        &\!\!\rho_c(t\0bf+(1-t)\widetilde{\Vbf}_c^{\text{lk}\circ}(\delta_c^{\text{w}}))\notag\\
        &\!\!\leq t\rho_c(\0bf)\!+\!(1\!-\!t)\rho_c(\widetilde{\Vbf}_c^{\text{lk}\circ}(\delta_c^{\text{w}}))\!=\!t\rho_c(\0bf)+\!(1\!-\!t)\delta_c^{\text{w}}.\!\!\label{EQ:conv2}
\end{align}

We further discuss (\ref{EQ:conv2}) in the following two subcases: 

2.\romannum{1}) If $\rho_c(\0bf)=\delta_c^{\text{w}}$, we have
\begin{align}
        \rho_c(t\mathbf{0}+(1-t)\widetilde{\mathbf{V}}_c^{\text{lk}\circ}(\delta_c^{\text{w}}))\le\delta_c^{\text{w}}<\delta_c.\label{EQ:conv3}
\end{align}
From (\ref{EQ:conv1}) and (\ref{EQ:conv3}), $(1-t)\widetilde{\mathbf{V}}_c^{\text{lk}\circ}(\delta_c^{\text{w}}),\forall{t}\in(0,1]$
satisfies the Slater's condition.

2.\romannum{2}) If $\rho_c(\0bf)>\delta_c^{\text{w}}$, we can set $t'=\frac{\delta_c'-\delta_c^{\text{w}}}{\rho_c(\0bf)-\delta_c^{\text{w}}}$, for any $\delta_c'\in(\delta_c^{\text{w}},\min\{\delta_c,\rho_c(\0bf)\})$ such that
\begin{align}
        \rho_c(t'\mathbf{0}+(1-t')\widetilde{\mathbf{V}}_c^{\text{lk}\circ}(\delta_c^{\text{w}}))\le\delta_c'<\delta_c.\label{EQ:conv4}
\end{align}
From (\ref{EQ:conv1}) and (\ref{EQ:conv4}), we have found a $t'\in(0,1)$ such that $(1-t')\widetilde{\mathbf{V}}_c^{\text{lk}\circ}(\delta_c^{\text{w}})$ satisfies the Slater's condition. 

Combining Cases 1) and 2), we complete the proof.\endIEEEproof

\section{Proof of Lemma \ref{LM:deltac}}\label{APP:deltac}

\textit{Proof:} Since the strong duality holds for $\Pc_c^{\text{w}}(\theta_c)$ for any $\theta_c\ge0$, we can solve $\Pc_c^{\text{w}}(\theta_c)$ through its dual problem. The Lagrangian for $\Pc_c^{\text{w}}(\theta_c)$ is
\begin{align}
        &L_c^{\text{w}}(\widetilde{\Vbf}_c,\lambda_c;\theta_c)\notag\\
        &=(1-\theta_c)f_c(\widetilde{\Vbf}_c)+\theta_c\rho_c(\widetilde{\Vbf}_c)+\lambda_c(\Vert\widetilde{\Vbf}_c\Vert_F^2-P_c^{\text{max}})\notag
\end{align}
where $\lambda_c\ge0$ is the Lagrange multiplier associated with constraint (\ref{EQ:MC_gc}). The dual problem of $\Pc_c^{\text{w}}(\theta_c)$
is given by
\begin{align*}
        \Dc_c^{\text{w}}(\theta_c):\quad\max_{\lambda_c\ge0}\min_{\widetilde{\Vbf}_c}\quad{L}_c^{\text{w}}(\widetilde{\Vbf}_c,\lambda_c;\theta_c).
\end{align*}
Let $(\widetilde{\Vbf}_c^{\text{w}\circ}(\theta_c),\lambda_c^\circ(\theta_c))$
denote an optimal solution to $\Dc_c^{\text{w}}(\theta_c)$. By setting $\theta_c=\frac{\nu_c^\circ(\delta_c)}{1+\nu_c^\circ(\delta_c)}\in[0,1)$ and adding a constant $-\frac{\nu_c^\circ(\delta_c)\delta_c}{1+\nu_c^\circ(\delta_c)}$
to the objective in $\Dc_c^{\text{w}}\left(\frac{\nu_c^\circ(\delta_c)}{1+\nu_c^\circ(\delta_c)}\right)$, the optimization problem is equivalent to
\begin{align}
        \max_{\tilde{\lambda}_c\ge0}~\min_{\widetilde{\Vbf}_c}\quad&{f}_c(\widetilde{\Vbf}_c)+\nu_c^\circ(\delta_c)[\rho_c(\widetilde{\Vbf}_c)-\delta_c]\notag\\
        &\quad+\tilde{\lambda}_c(\Vert\widetilde{\Vbf}_c\Vert_F^2-P_c^{\text{max}})\label{EQ:LM3-1}
\end{align}
where $\tilde{\lambda}_c\triangleq\lambda_c(1+\nu_c^\circ(\delta_c))$.
The dual problem $\Dc_c^{\text{lk}}(\delta_c)$ for any $\delta_c>\delta_c^{\text{w}}$ is given by
\begin{align}
       \max_{\nu_c\ge 0,\mu_c\ge0}~\min_{\widetilde{\Vbf}_c}\quad&{f}_c(\widetilde{\Vbf}_c)+\nu_c[\rho_c(\widetilde{\Vbf}_c)-\delta_c]\notag\\
        &\quad+\mu_c(\Vert\widetilde{\Vbf}_c\Vert_F^2-P_c^{\text{max}}).\label{EQ:LM3-2}
\end{align}
Comparing $\Dc_c^{\text{w}}\left(\frac{\nu_c^\circ(\delta_c)}{1+\nu_c^\circ(\delta_c)}\right)$ in (\ref{EQ:LM3-1}) with $\Dc_c^{\text{lk}}(\delta_c)$ in (\ref{EQ:LM3-2}), we can treat $\nu_c^\circ(\delta_c)$ in $\Dc_c^{\text{w}}\left(\frac{\nu_c^\circ(\delta_c)}{1+\nu_c^\circ(\delta_c)}\right)$ as a predetermined value of Lagrange multiplier $\nu_c$ in $\Dc_c^{\text{lk}}(\delta_c)$. Noting that $\nu_c^\circ(\delta_c)$ is optimal for $\Dc_c^{\text{lk}}(\delta_c)$, we have $\widetilde{\Vbf}_c^{\text{w}\circ}\left(\frac{\nu_c^\circ(\delta_c)}{1+\nu_c^\circ(\delta_c)}\right)\in\Vc_c^{\text{lk}}(\delta_c)$ and $\widetilde{\Vbf}_c^{\text{lk}\circ}(\delta_c)\in\Vc_c^{\text{w}\circ}\left(\frac{\nu_c^\circ(\delta_c)}{1+\nu_c^\circ(\delta_c)}\right)$. Thus, we complete the proof.
\endIEEEproof

\section{Proof of Theorem \ref{Thm:LK}}\label{APP:LK}

\textit{Proof:} By Lemma~\ref{LM:fea}, we only consider $\delta_c\ge\delta_c^{\text{w}}$ for feasible $\Pc^{\text{lk}}_c(\delta_c)$. By Lemma \ref{LM:deltac}, claim \romannum{1}) holds. We now prove claim \romannum{2}). From Lemma~\ref{LM:fea}, $\Pc^{\text{lk}}_c(\delta_c^{\text{w}})$ is feasible, we have  $\rho_c(\widetilde{\mathbf{V}}_c^{\text{lk}\circ}(\delta_c^{\text{w}}))=\delta_c^{\text{w}}$ and $\widetilde{\Vbf}_c^{\text{lk}\circ}(\delta_c^{\text{w}})$ satisfies (\ref{EQ:MC_gc}). Thus, $\widetilde{\Vbf}_c^{\text{lk}\circ}(\delta_c^{\text{w}})$ is also a feasible solution to $\Pc_c^{\text{w}}(1)$. Noting that $\rho_c(\widetilde{\Vbf}_c^{\text{w}\circ}(1))=\delta_c^{\text{w}}$
in (\ref{EQ:defdeltac-}), we have $\rho_c(\widetilde{\Vbf}_c^{\text{lk}\circ}(\delta_c^{\text{w}}))=\rho_c(\widetilde{\Vbf}_c^{\text{w}\circ}(1))$. Since $\rho_c(\widetilde{\Vbf}_c^{\text{w}\circ}(1))$ is the minimum objective value of  $\Pc_c^{\text{w}}(1)$, $\widetilde{\Vbf}_c^{\text{lk}\circ}(\delta_c^{\text{w}})$ is also an optimal solution to $\Pc_c^{\text{w}}(1)$, \ie $\widetilde{\Vbf}_c^{\text{lk}\circ}(\delta_c^{\text{w}})\in\mathcal{V}_c^{\text{w}}(1)$, for any $\widetilde{\Vbf}_c^{\text{lk}\circ}(\delta_c^{\text{w}})\in\mathcal{V}_c^{\text{lk}}(\delta_c^{\text{w}})$. Therefore, we complete the proof.\endIEEEproof

\bibliographystyle{IEEEtran}
\bibliography{CoMPWNV}

\end{document}